\documentclass[preprint,12pt]{elsarticle}

\usepackage{amssymb}
\usepackage{rotating}
\usepackage{siunitx}

\newcommand {\apgt} {\ {\raise-.5ex\hbox{$\buildrel>\over\sim$}}\ }
\newcommand {\aplt} {\ {\raise-.5ex\hbox{$\buildrel<\over\sim$}}\ }

\journal{Journal of Computational Science}

\begin{document}

\begin{frontmatter}



\title{Analysing and Modelling the Performance of the HemeLB Lattice-Boltzmann Simulation Environment}


\author[UCL]{Derek Groen}
\author[UCL]{James Hetherington}
\author[UCL]{Hywel B. Carver}
\author[UCL]{Rupert W. Nash}
\author[UCL]{Miguel O. Bernabeu}
\author[UCL]{Peter V. Coveney}

\address{Centre for Computational Science, University College London,
London, United Kingdom, e-mail: d.groen@ucl.ac.uk}

\begin{abstract}
We investigate the performance of the HemeLB lattice-Boltzmann simulator for
cerebrovascular blood flow, aimed at providing timely and clinically relevant
assistance to neurosurgeons. HemeLB is optimised for sparse geometries,
supports interactive use, and scales well to 32,768 cores for problems with
$\sim$81 million lattice sites. We obtain a maximum performance of 29.5 billion
site updates per second, with only an 11\% slowdown for highly sparse
problems (5\% fluid fraction). We present steering and visualisation
performance measurements and provide a model which allows users to predict the
performance, thereby determining how to run simulations with maximum accuracy
within time constraints.
\end{abstract}


\end{frontmatter}

\section{Introduction}

Recent progress in imaging and computing technologies has resulted in an
increased adoption of computational methods in the life sciences. Using modern
imaging methods, we are now able to scan the geometry of individual vessels
within patients and map out potential sites for vascular malformations such as
intracranial aneurysms.  Likewise, recent increases in computational capacity
and algorithmic improvements in simulation environments allow us to simulate
blood flow in great detail. The HemeLB lattice-Boltzmann
application~\cite{Mazzeo:2008} aims to combine these two developments, thereby
allowing medical scans to be used as input for blood flow simulations. It also
enables clinicians to run such simulations in real-time, providing runtime
visualisation feedback as well as the ability to steer the simulation and its
visualisation~\cite{Mazzeo:2010}. One principal long-term goal for HemeLB is to
act as a production toolkit that provides both timely and clinically relevant
assistance to surgeons. To achieve this we must not only perform extensive
validation and testing for accuracy, reliability, usability and performance,
but also ensure that the legal environment and the medical and computational
infrastructure are made ready for such use cases~\cite{Sadiq:2008}. 

In this work we investigate the performance aspects of the HemeLB environment,
taking into account the core lattice-Boltzmann (LB) simulation code and the
visualisation and steering facilities. We present performance measurements from
a large number of runs using both sparse and non-sparse geometries and the
overheads introduced by visualisation and steering. Medical doctors treating
patients with intracranial aneurysms are frequently confronted with very short
time scales for decision-making. For HemeLB to be useful in such environments,
it is therefore not only essential that the code simulates close to real-time,
but also that the length of a simulation can be reliably predicted in advance.
We demonstrate that it is possible to accurately characterise CPU and network
performance at low core counts and integrate this information into a model that
predicts performance for arbitrary problem sizes and core counts.

\subsection{Overview of HemeLB}

HemeLB is a massively parallel lattice-Boltzmann simulation framework that
allows interactive use, eventually in a medical environment. Segmented angiographic data
from patients can be read in by the HemeLB Setup Tool, which allows the user to
indicate the geometric domain to be simulated using a graphical user interface.
The geometry is then discretised into a regular grid, which is used to run
HemeLB simulations. The core HemeLB code, written in C++, consists of a
parallelised lattice-Boltzmann application which is optimised for sparse
geometries such as vascular networks by use of indirect addressing. We
precompute the addresses of neighbouring points within a single one-dimensional
array instead of requiring that the points be stored in a dense,
three-dimensional array. HemeLB also constructs a load-balanced domain
decomposition at runtime, allowing the user to run simulations at varying core
counts with the same simulation domain data.  HemeLB is highly scalable due to
a well-optimised communication strategy and the locality of interactions and
communications in the parallelised lattice-Boltzmann algorithm. The File I/O 
operations are done in parallel using MPI-IO by a group of {\em reading 
processes}, which can be adjusted in size using a compile-time parameter. 

The HemeLB Steering Client is a light-weight tool that allows users to connect
remotely to their HemeLB simulation, receive real-time visual feedback and
modify parameters of the simulation at runtime. Here, the visualisations are
generated on-site within HemeLB, using a hand-written ray-tracing
kernel~\cite{Mazzeo:2010}.  In our work we run HemeLB with the steering server
code enabled. As a result, one core is reserved for steering purposes, whether
or not a client is connected, and is thereby excluded from the LB calculations.

HemeLB relies on ParMETIS version 4.0.2~\cite{parmetis-site} to perform its
domain decomposition.  It constructs an initial guess using a basic graph
growing partitioning algorithm (see Mazzeo et al.~\cite{Mazzeo:2008} for details), which it then passes to
ParMETIS for optimisation using the {\tt ParMETIS\_V3\_PartKway()} function.
Constructing the initial guess requires less than a second of runtime in all
cases, but the ParMETIS optimisation typically adds between 5 and 30 seconds to
the initialisation time. We discuss several technical aspects and performance
implications of our decomposition routine in Section~\ref{Sect:extract}.

HemeLB uses a coalesced asynchronous communication strategy to optimize its
scalability~\cite{Carver:2012-2}. This system bundles all communications for
each iteration (e.g., exchanges required for the LB algorithm, steering and
visualisations) into a single batch of non-blocking communication messages, one
for each data exchange of non-zero size between a pair of processes in each
direction. As a result, each iteration of HemeLB's core loop has only one {\tt
MPI\_Wait} synchronisation point, minimising the latency overhead of HemeLB
simulations. Communication of variable length data is spread over two
iterations, the sizes being transferred during the first iteration while the
actual exchange takes place during the second one. 

The coalesced communication system is also used for the phased broadcast and
reduce operations which are required for the visualisation and steering
functionality.  Here HemeLB arranges the processes into an $n$-tree and, for
broadcasts, sends data from one level of the tree to the level below over
successive iterations. For reductions, data is sent up one level of the tree
over successive iterations.  Hence, both operations can take $O(\log (p))$
iterations, for $p$ cores.  In this approach HemeLB does require some
additional memory for communication buffers. Additionally, the responsiveness
of the steering is constrained, as data arriving in the top-most node takes
$O(\log (p))$ iterations to be spread to all nodes.

\subsection{Related work}

A large num)ber of researchers have investigated the performance aspects of
various LB simulation codes over the past decade. These investigations have
been done without real-time visualisation or steering enabled, and frequently
use non-sparse geometries. We present a performance analysis of both sparse
geometries and interactive usage modes in this work.  Pohl {\em et
al.}~\cite{Pohl:2004} compared the performance of LB codes across three
supercomputer architectures, and concluded that the network and memory
performance (bandwidth and latency) are dominant components in establishing a
high LB calculation performance. Geller {\em et al.}~\cite{Geller:2006}
compared the performance of an LB code with that of several finite element and
finite volume solvers, and deduced that LB offers superior efficiency in flow
problems with small Mach numbers. Williams et al.~\cite{Williams:2011} presented a 
hierarchical autotuning model for parallel lattice-Boltzmann, and report a performance
increase of more than a factor 3 in their simulations. Several groups have considered the
performance of LB solvers on general-purpose graphics processing unit (GPGPU)
architectures. In these studies, they introduced a number of improvements, such
as non-uniform grids~\cite{Schonherr:2011}, more efficient memory management
strategies~\cite{Habich:2012,Obrecht:2011} and LB codes which run across
multiple GPUs~\cite{Obrecht:2011-2,Myre:2011,Gray:2011}. Other performance investigations 
include a comparison between different LB implementations~\cite{Donath:2008}, 
hybrid parallelisations for multi-core architectures in
general~\cite{Heuveline:2009,Schonherr:2011,Fietz:2012} and performance analysis of LB
codes on Cell processors~\cite{Harvey:2008,Williams:2009,Biferale:2010}.

A few studies within the physiological domain are of special relevance to this work.
These include a performance analysis of a blood-flow LB solver using a range of
sparse and non-sparse geometries~\cite{Bernsdorff:2008} and a performance
prediction model for lattice-Boltzmann solvers~\cite{Axner:2007,Axner:2008}. This
performance prediction model can be applied largely to our HemeLB application,
although HemeLB uses a different decomposition technique and performs real-time
rendering and visualisation tasks during the LB simulations. Mazzeo and
Coveney~\cite{Mazzeo:2008} studied the scalability of an earlier version of
HemeLB. However, the current performance characteristics of HemeLB are
substantially enhanced due to numerous subsequent advances in the code, amongst
others: an improved hierarchical, compressed file format; the use of ParMETIS
to ensure good load-balance; the coalesced communication patterns to reduce
the overhead of rendering; use of compile-time polymorphism to avoid virtual
function calls in inner loops.

\section{Performance analysis}

\begin{figure}[!t] \centering
\includegraphics[width=5.0in]{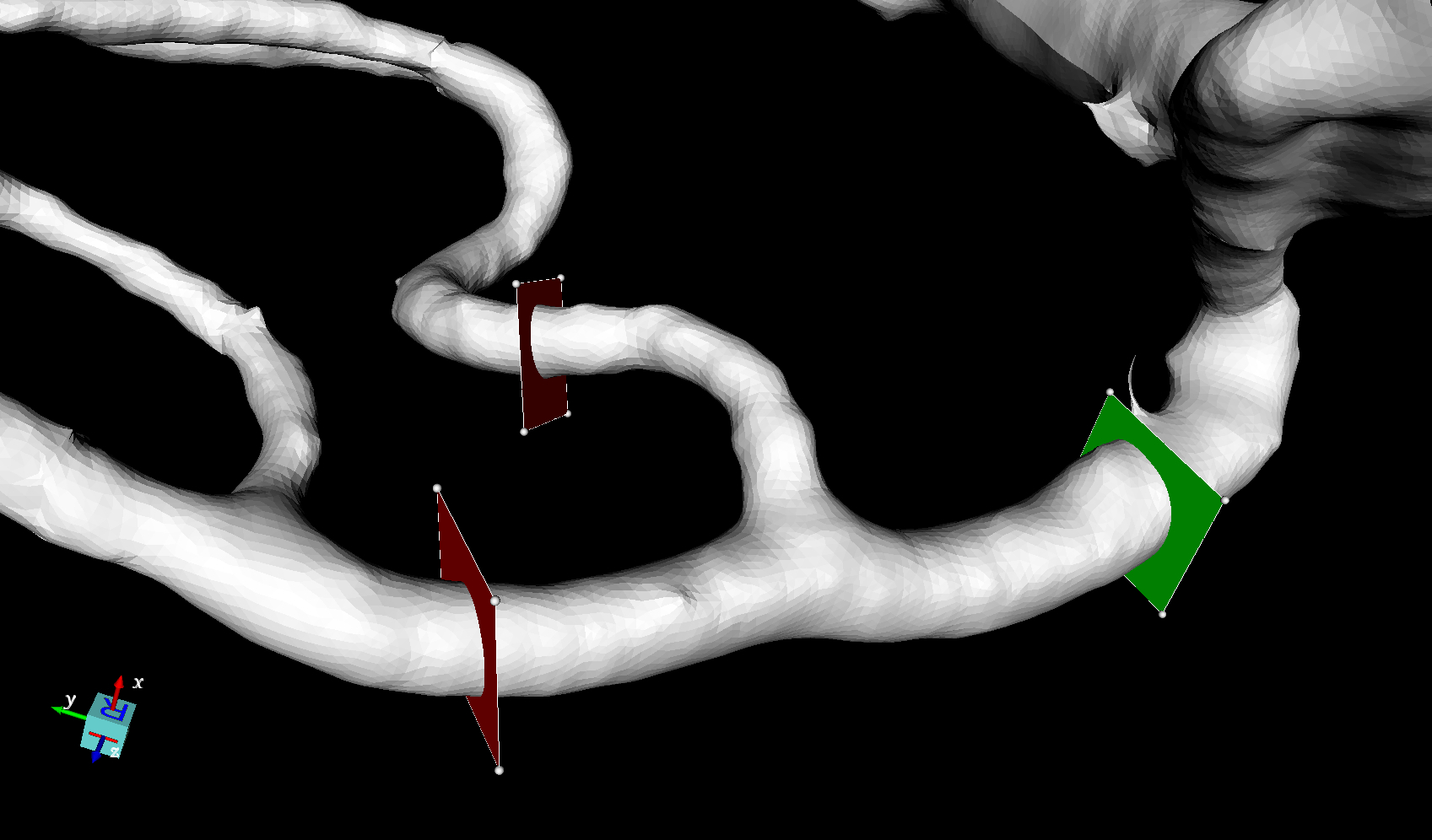} \caption{Graphical
overview of the bifurcation geometry in the HemeLB Setup Tool. We used this
geometry to generate the Bifurcation and Large Bifurcation simulation domains.
Inlets are shown by green planes, outlets by red planes.}
\label{Fig:bifurcation} \end{figure}

\begin{figure}[!t] \centering
\includegraphics[width=5.0in]{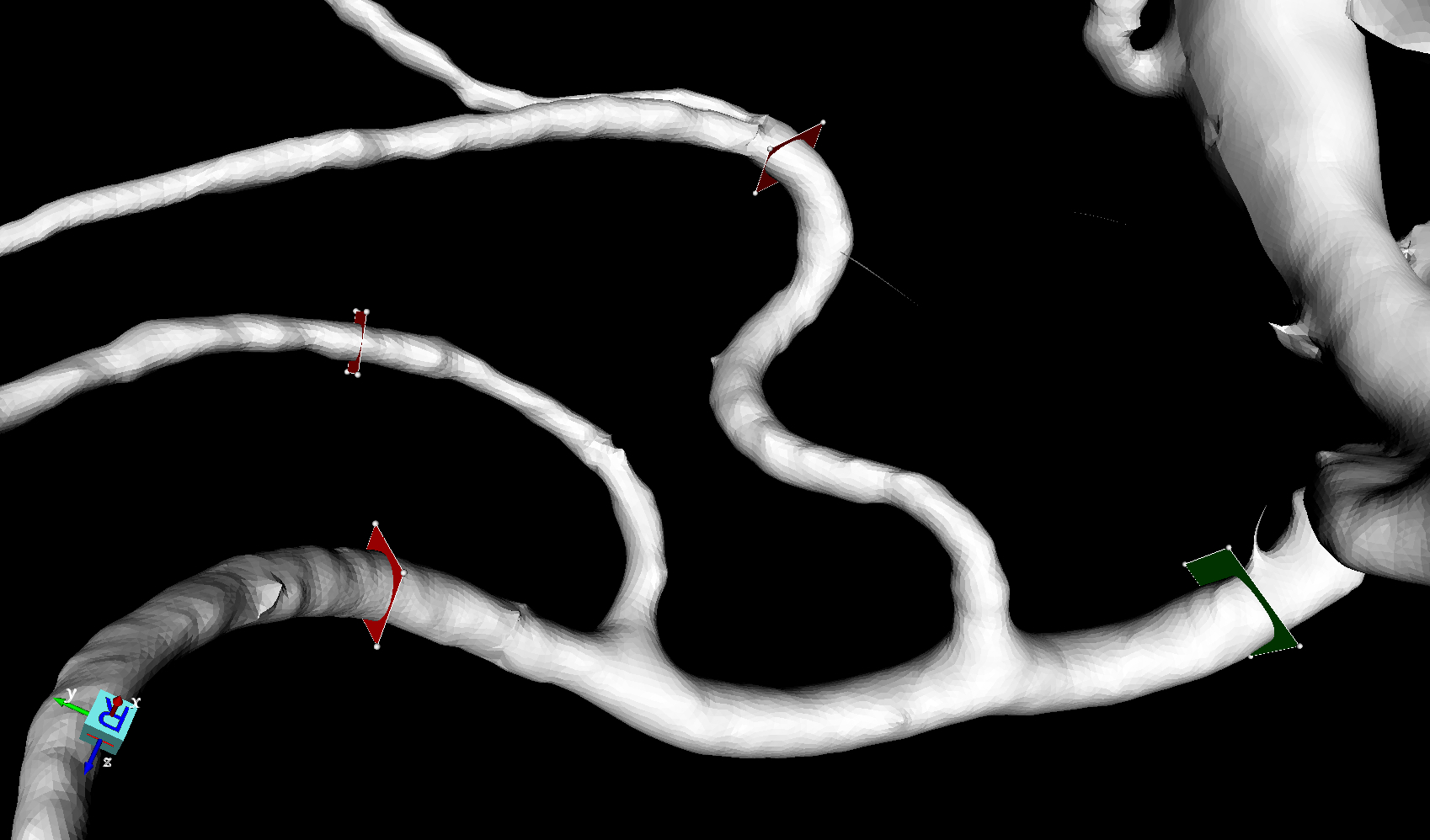} \caption{Graphical
overview of the network geometry in the HemeLB Setup Tool. We used this
geometry to generate the Network, Large Network and Small Network simulation
domains.} \label{Fig:network} \end{figure}

%\begin{figure}[!t] \centering
%\includegraphics[width=3.3in]{cylinder-setuptool.jpg} \caption{Graphical
%overview of the cylinder geometry in the HemeLB Setup Tool. The thickness of
%the cylinder is one-tenth of its diameter.} \label{Fig:cylinder} \end{figure}

We benchmarked HemeLB using simulation domains based on three distinct
geometries, a vascular network (see Fig.~\ref{Fig:network}, used to generate
three simulation domains), a bifurcation of vessels (see
Fig.~\ref{Fig:bifurcation}, used to generate two simulation domains) and a
cylinder. Both the network and the bifurcation geometries are sections of an
intracranial vasculature model that has been constructed from multiple
rotational angiography scans of a patient with an intracranial aneurysm treated
at the U.K. National Hospital for Neurology and Neurosurgery. The third and
least sparse geometry is an artificially created cylinder.  We present an
overview of the simulation domains we generated and use in our runs in
Table~\ref{Tab:ICs}. We also provide a brief description of the sparseness of
each generated simulation domain. Our runs were impulsively started, 
applying a pressure gradient across the simulation domain, using Nash in-outlet
conditions (Nash et al., in preparation).

\begin{table}[!t] \caption{Overview of the simulation domains used in our
experiments. The percentage of the simulated box that consists of active fluid
sites is given by the fluid fraction.  Non-active fluid sites do not count
towards the number of lattice sites in the simulation.} \label{Tab:ICs}
\centering 
\begin{tabular}{|l|l|l|} 
\hline 
Name & \# of lattice sites & fluid fraction\\ 
\hline 
Bifurcation       & 19,808,107 & 11\% \\ 
Cylinder          & 15,607,040 & 65\%  \\ 
Network           & 18,836,545 & 5.1\% \\ 
Large Bifurcation & 81,132,544 & 11\%  \\ 
Large Network     & 44,650,496 & 5.1\% \\
Small Network     & 77,182     & 5.1\% \\ 
\hline
\end{tabular} 
\end{table}

\subsection{Performance of LB computations}\label{Sect:coreLBperf}

\begin{figure}[!t]
\centering
\includegraphics[width=5.0in]{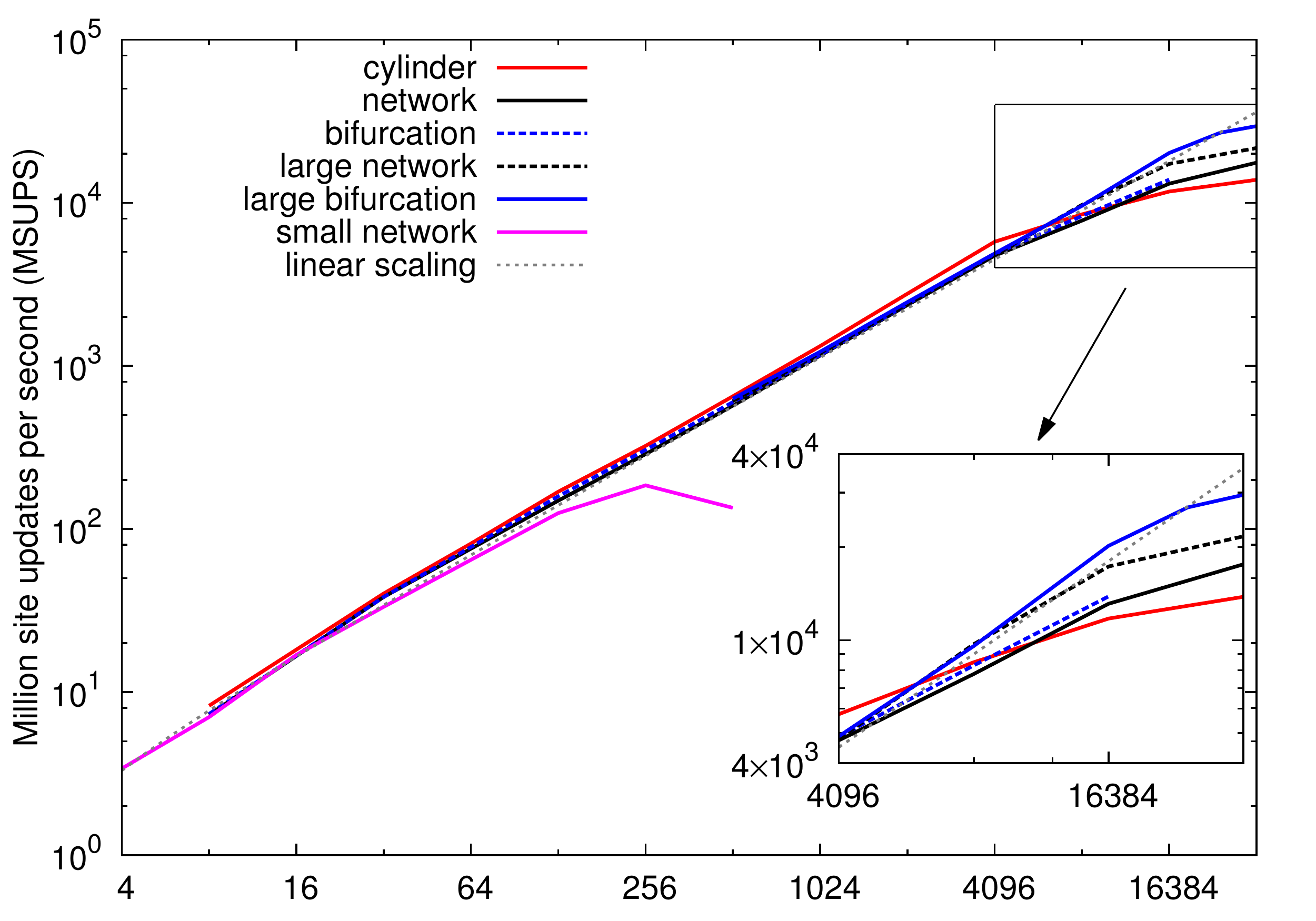}
\caption{Lattice site updates per second (SUPS) as a function of the number 
of cores used for simulations run on the HECToR Cray XE6 machine. We run
simulations using each of the six simulation domains (Cylinder, Network, 
Bifurcation, Large Bifurcation, Large Network and Small Network).}
\label{Fig:Scalability-sups}
\end{figure}

\begin{figure}[!t]
\centering
\includegraphics[width=5.0in]{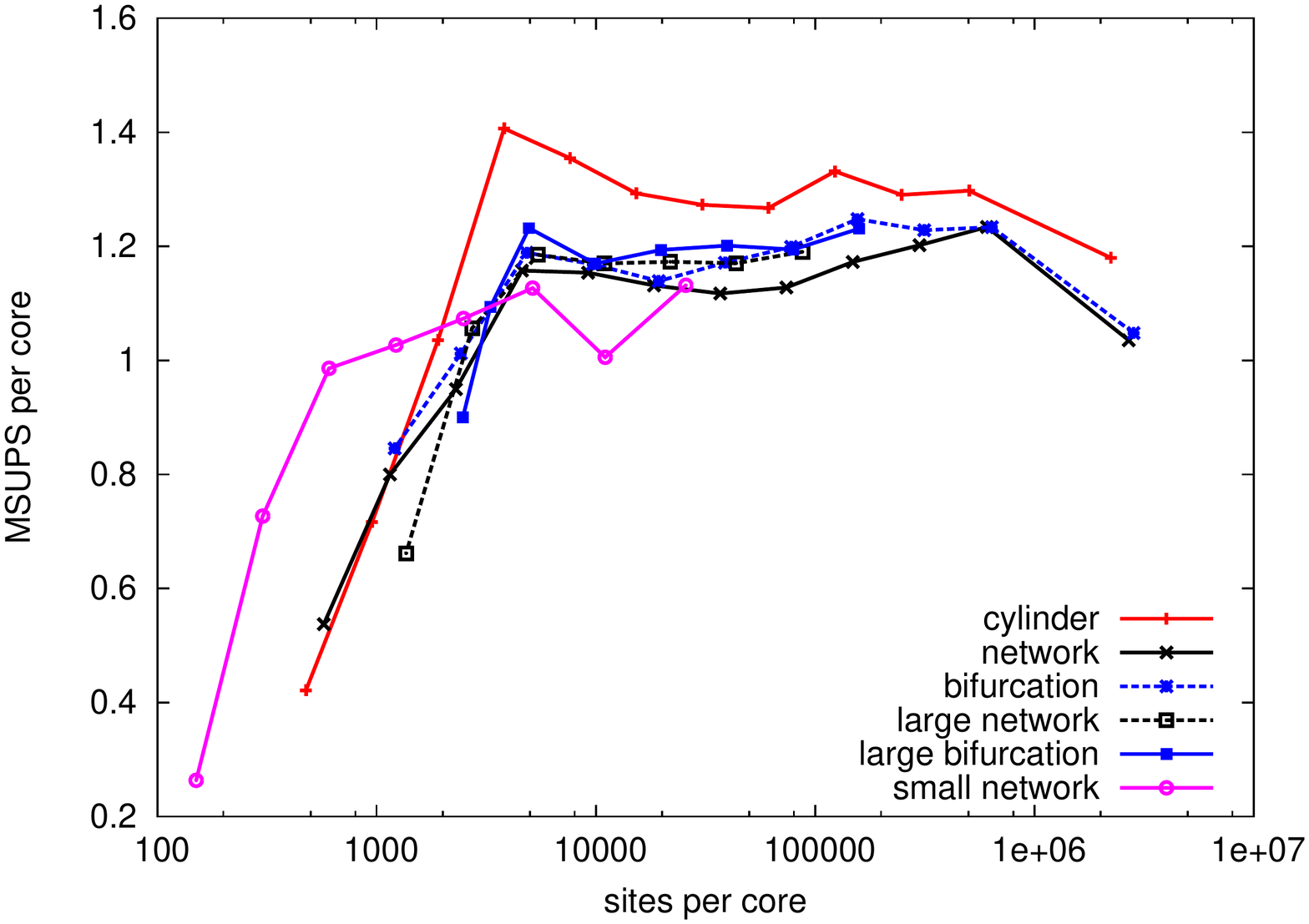}
\caption{Site updates per second (SUPS) per core averaged over all cores used in the simulation
(excluding the one used for steering) as a function of the number of sites per core, for six LB 
problems.}
\label{Fig:Scalability-universal}
\end{figure}

We have run blood flow simulations using the simulation domains listed in
Table~\ref{Tab:ICs} using up to 32,768 cores on the HECToR Phase 3
supercomputer at EPCC in Edinburgh, United Kingdom. The HECToR machine is a
Cray XE6 with 90,112 cores (2.3GHz AMD Opteron 6276), and has a peak
performance of 9.2 GFLOP/s per core. Our simulations were done using a
15-directional lattice-Boltzmann kernel (D3Q15), the Lattice
Bhatnagar-Gross-Krook~\cite{Bhatnagar:1954} model with simple bounce-back boundary conditions and a
fixed physical viscosity of \SI{0.004}{\pascal\second}. We present the scalability
results for all simulation domains in Figure~\ref{Fig:Scalability-sups}. We
find that the small network simulation domain scales near-linearly up to 128
cores, despite consisting of only 77,182 lattice sites. All of the medium-sized
simulation domains (Bifurcation, Cylinder and Network) scale linearly to 8,192
cores. However, the communication overhead and load imbalance reduce the
performance on higher core counts. The two largest simulation domains (Large
Bifurcation and Large Network) show linear scaling from 512 cores up to 16,384 
cores, and significant speedup to 32,768 cores, achieving a maximum performance of 29.5
billion site updates per second (SUPS). The performance obtained at 8,196 cores
for the medium-sized bifurcation corresponds to 419 timesteps per second, or
646 times slower than real-time for a maximum timestep as limited by incompressibility
constraints. The maximum timestep here is estimated by the need to keep the Mach number
below 0.05, using a typical blood velocity for vessels of this size of 25 cm/s.
At this rate, it takes HemeLB 553 seconds to simulate one heartbeat with a
resolution of around 100 lattice points across a vessel diameter. We present
the performance in SUPS per core as a function of the number of sites per core
in Figure~\ref{Fig:Scalability-universal}, demonstrating that the SUPS per core
is largely independent of other factors.

\subsection{Visualisation performance}\label{Sect:vizperf}

\begin{figure}[!t]
\centering
\includegraphics[width=5.0in]{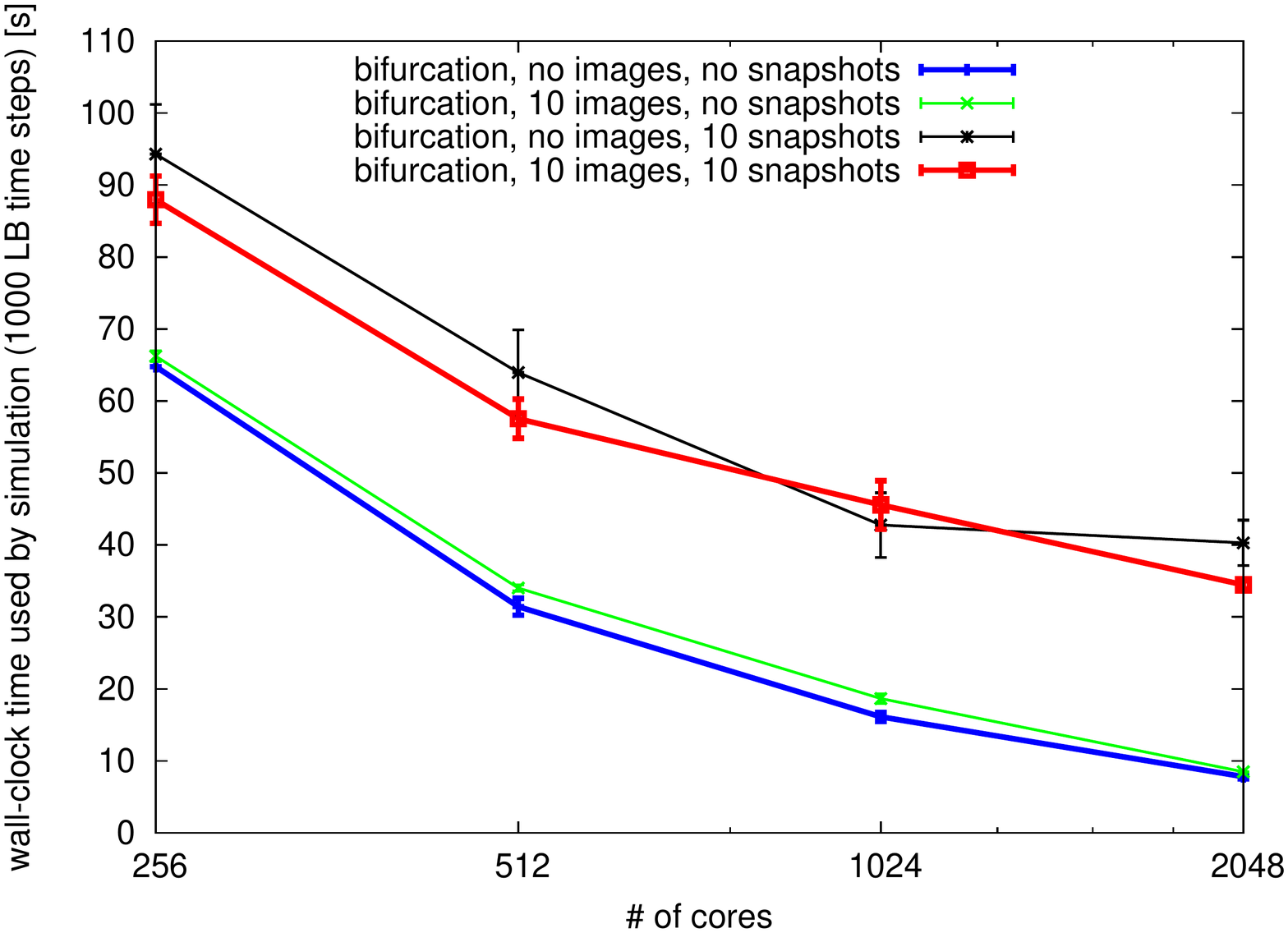}
\caption{Time spent to simulate 100 time steps as a function of the number of 
cores used for four settings: no snapshots and no images (blue), images only 
(green), snapshots only (black) and snapshots and images (red). We averaged the 
results from runs including any form of snapshot or image writing over three 
executions, and included a standard deviation error bar with each data point.}
\label{Fig:Scalability-vis}
\end{figure}

One of the features that sets HemeLB apart from many other LB codes is its
ability to perform {\em in situ} rendering of the geometry at
runtime~\cite{Mazzeo:2010}, using a parallelised ray-tracing algorithm. The
communication needs of the ray-tracing algorithm have been combined with those
of the main simulation algorithm, through the coalesced communication strategy,
massively improving the scaling when rendering frames. The images rendered by
HemeLB can either be stored on disk for future reference or they can be
forwarded as a streaming visualisation to the steering client. In this section
we present several simulations where we assess the overhead introduced by
rendering images, as well as that introduced by writing snapshots of the
simulation data. These snapshots store the hydrodynamic variables at each
lattice point, recording all information of physical relevance which is useful
for visualisation and post-processing. File I/O operations are done in HemeLB
using a subset of all processes, the {\em reading group}. Within this work, we
adopted a reading group size of 32 processes, or the number of processes used
by HemeLB, whichever was smaller.  We have run four types of simulations using
the Bifurcation simulation domain, one with snapshots and image-rendering
disabled, one where we write snapshots to disk (10 snapshots per 1000 time
steps, with each snapshot being 604MB in size), one where we render and write
images to disk (10 rendered images per 1000 time steps, with each image being
180kB in size) and one with both snapshots and images enabled. We have carried
out the tests using 256, 512, 1024 and 2048 cores. We present our results in
Fig.~\ref{Fig:Scalability-vis}.  Here the overhead for rendering and writing
images is marginal, and adds no more than a few percent to the execution time
in most cases. Simulations which have snapshot writing enabled are both
considerably slower and have more variable performance, due to the high disk
activity involved with snapshot writing. When snapshot writing is enabled, the
overhead caused by image rendering is difficult to observe, as the standard
deviation bars of the performance measurements with and without images overlap.
When the simulation writes 10 snapshots over 1000 LB steps, we observe an
increase in the wall-clock time of $\sim$24 seconds.

\begin{figure}[!t]
\centering
\includegraphics[width=5.0in]{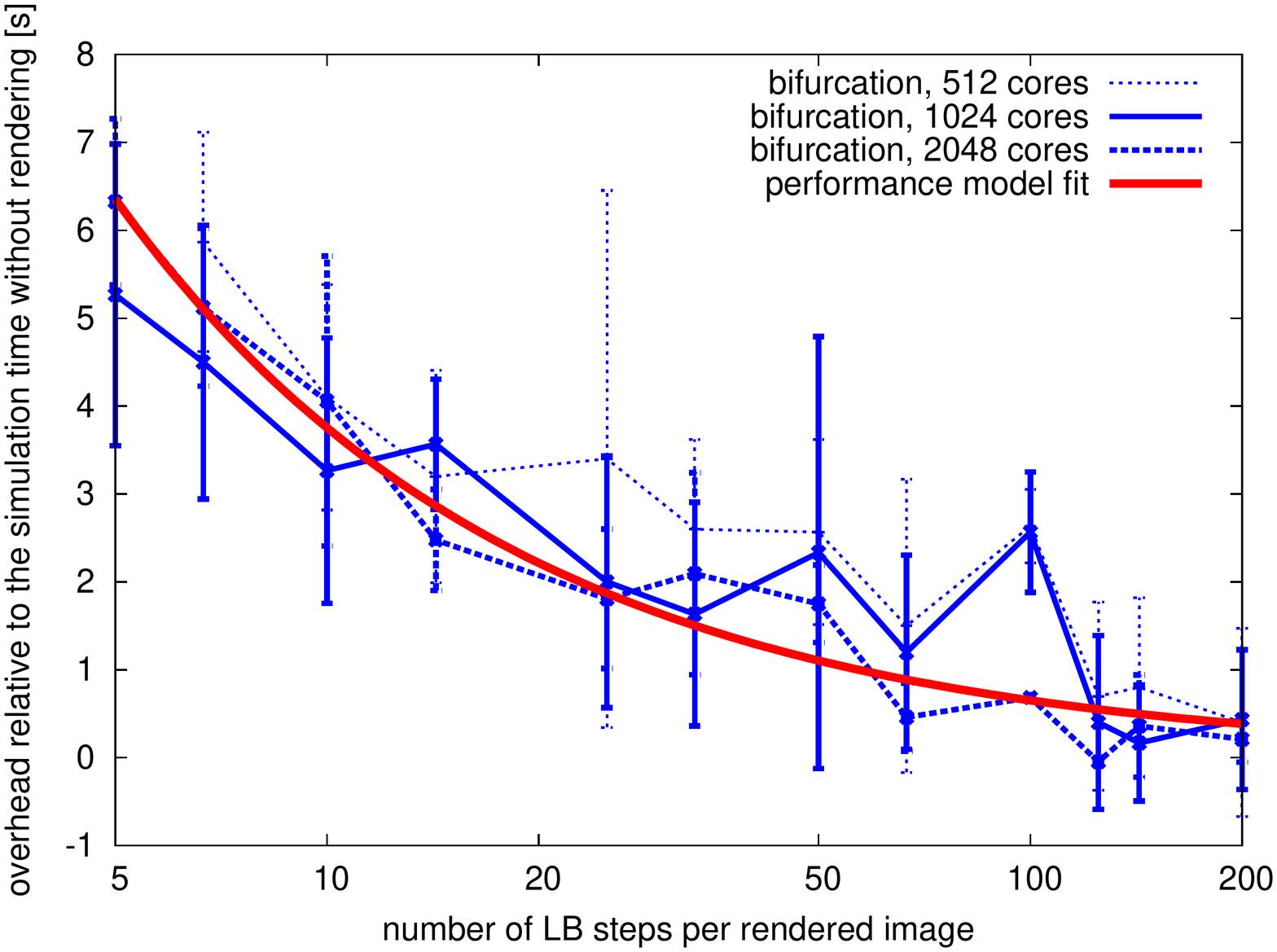}
\caption{Overhead in seconds relative to the simulation time without images 
rendered as a function of the number of LB steps per image rendered and written. 
The simulation with 0 images rendered took 31.4, 16.1 and 7.81 seconds on 
respectively 512, 1024 and 2048 cores. We averaged the measurement of the runs 
over three executions. Error bars are the resulting standard deviations. The 
prediction of our performance model, presented in 
Section~\ref{Sect:vis-model}, is given by the thick solid red curve.}
\label{Fig:Scalability-vis-render}
\end{figure}

We have also run several simulations of 1000 LB steps where we render and write
an image to disk every 5 to 200 LB steps. The results for these runs (which
were done using 1024 and 2048 cores) are given in
Fig.~\ref{Fig:Scalability-vis-render}.  Without rendering the simulations took
31.4, 16.1 and 7.81 seconds on 512, 1024 and 2048 cores respectively. We
observe an overhead of less than two seconds per 1000 LB steps if we render and
write no more than 10 images during that period. However, the performance
deteriorates somewhat when we write more images, with a maximum measured
overhead of $\sim$6.5 seconds.  We also again observe some jitter in our
results, for example in the 1024 core simulation that rendered one image every
50 steps, which we attribute to fluctuations in the file system performance of
the machine. Rendering one image per 5 LB steps using 2048 cores corresponds to
a frame rate of about 13.6 frames per second, more than sufficient for smooth
visualisations of the simulations in real time.

\subsection{Steering performance}

\begin{figure}[!t]
\centering
\includegraphics[width=5.0in]{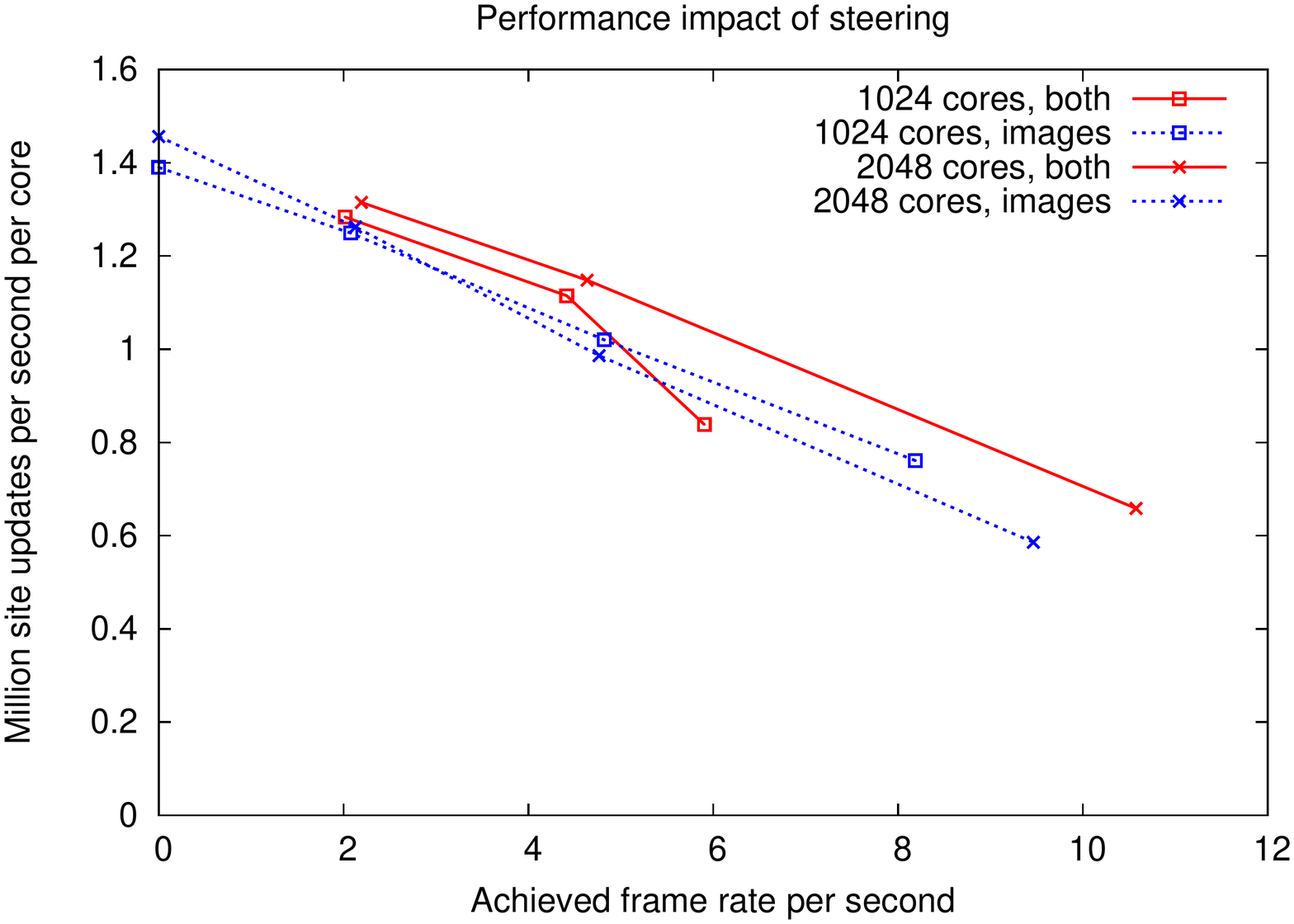}
\caption{Performance impact of running HemeLB with a connected steering client.
We show results for 1024 and 2048 cores without steering client (plotted at 
frame-rate zero), with the client used only for image streaming (images) and with
the client used both for image streaming and steering the HemeLB simulation (both).}
\label{Fig:Steering}
\end{figure}

The previous subsection isolates the performance impact of the visualisation
and rendering, with images written to disk. Here we study the performance
impact of the HemeLB steering component, using the Cylinder simulation domain,
where images are streamed over the network to a client. In this case, HemeLB
produces images as described in Section~\ref{Sect:vizperf}, optionally limited
by a maximum frame-rate per second. We also look at the performance impact of
sending steering messages from the client to the HemeLB steering component.  In
order to obtain reproducible data, the steering client is set up with a
scripted set of simulated user actions (orbiting the view point for image
rendering). These results are presented in Table~\ref{Tab:SteeringResults} and
in Figure~\ref{Fig:Steering} and were produced with the steering client running
on the HECToR login node. For a frame-rate of 4.6 frames per second, which is
usable for scientific steering, with bidirectional communication between client
and server, corresponding to 32 LB steps per rendered image, we observe an
overhead of 28\%. 

\begin{table}[!t]
\caption{Performance impact of running HemeLB with a connected steering client,
simulating the Cylinder simulation domain using 1024 and 2048 cores. Here the mode is the method of running HemeLB, which can be without 
client (none), with the client used only for image streaming (images) or with
the client used both for image streaming and steering the HemeLB simulation (both).}
\label{Tab:SteeringResults}
\centering
\begin{tabular}{|llllll|}
\hline
$p$ & mode &\multicolumn{2}{|c|}{frame-rate (1/s)} & MSUPS    & mean LB steps  \\
    & & requested & achieved                 & per core & per image     \\
\hline
1024 & none   & -    & -    & 1.39 & -    \\
1024 & both   & 2.0  &  2.0 & 1.28 & 41.5 \\
1024 & images & 2.0  &  2.1 & 1.25 & 39.3 \\
1024 & both   & 5.0  &  4.4 & 1.11 & 16.5 \\
1024 & images & 5.0  &  4.8 & 1.02 & 13.8 \\
1024 & both   & max  &  5.9 & 0.84 & 11.3 \\
1024 & images & max  &  8.2 & 0.76 & 6.0  \\
2048 & none   & -    & -    & 1.46 & -    \\
2048 & images & 2.0  &  2.1 & 1.26 & 77.2 \\
2048 & both   & 2.0  &  2.2 & 1.32 & 78.6 \\
2048 & both   & 5.0  &  4.6 & 1.15 & 32.2 \\
2048 & images & 5.0  &  4.8 & 0.99 & 26.9 \\
2048 & images & max  &  9.5 & 0.59 & 8.0  \\
2048 & both   & max  & 10.6 & 0.66 & 8.1  \\
\hline
\end{tabular}
\end{table}

\subsection{Performance comparison with other codes}\label{Sect:perfcomp}

In this section we compare the performance of HemeLB with performance
measurements of other LB codes as found in the literature. We gathered the
number of million lattice site updates per second (MSUPS), the standard measure
of LB performance, reported for other implementations. HemeLB is strongly
optimised for efficiently handling sparse geometries while most codes are not,
making like-for-like comparison difficult. The other applications may not be
capable of simulating even moderate complexity domains, such as a cylinder, at
all or only at the cost of allocating memory to non-fluid sites. Additionally,
the directional resolution affects the number of calculations and memory
accesses required per site update, as well as the presence of other special
features, such as the additional presence of a D3Q15 magnetic field distribution
model in LBMHD~\cite{Williams:2011}. One particular example is 
LB3D~\cite{Harting:2004} version 7, which calculates a number of additional 
forces, and is strongly optimised for multi-phase flow 
at the expense of single-phase flow performance. For LB3D we therefore included
measurements for both single-phase and multi-phase flow performance.

We provide the LB performance configurations and results for
several well-known LB codes in Tables~\ref{Tab:CodeComparison} and
\ref{Tab:CodeComparison2}. The MSUPS per core results here are obtained by
dividing the total number of lattice site updates by the product of time spent
on LB iterations and the number of cores.  From each literature source, we
picked the result from the run that showed the best MSUPS per core while
running on at least one full processor. In the case of HemeLB we picked the
best result from the non-sparse Cylinder, as well as from the very sparse
Network and Large Network simulation domains, which are the only measurements
in the tables using sparse geometries.

When we examine bulk flow only, the MSUPS per core performance of HemeLB is
comparable with that achieved with LBMHD (although LBMHD calculates in 27
directions and HemeLB in 15), and about half of that achieved with Palabos on
similar AMD Opteron architectures. The performance of HemeLB, however, is
almost entirely preserved when using a very sparse simulation domain as HemeLB
does not allocate memory or computational effort for non-active lattice sites,
which are by definition common in sparse geometries. LBMHD has no known
optimisations for sparse geometries while Palabos features a partial
optimisation using the multi-block method~\cite{Chopard:2010}, of which we found
no performance data using sparse geometries in the literature. The multi-block
method is relatively inefficient because it allocates memory to some of the
non-fluid sites and uses data structures that grow in complexity when
off-lattice geometries are modelled more accurately. When a code is not
designed for sparse geometries, additional optimisations (e.g., cache
lookahead) are simpler to implement, hence the performance of a code which
supports sparse geometries may not match that of codes which exploit such
optimisations. Many of the benchmarks for other LB codes were performed on
non-Opteron architectures, making it difficult if not impossible to do a
one-on-one comparison. We nevertheless include these results for reference in
the lower part of Table~\ref{Tab:CodeComparison}.

\begin{table*}[!t]
\caption{Technical specifications of 12 LB simulations in our code comparison. 
We provide the name of the LB application used in the first column (including 
the source), followed by respectively the architecture used for the simulations 
and the number of cores used for the run.}
\label{Tab:CodeComparison}
\centering
\begin{tabular}{|lll|}
\hline
Name & Architecture & cores \\
     & (peak GFLOPS/core) & \\
\hline
HemeLB (Cylinder)             & AMD Opteron 6276 (9.2)   & 4096    \\
HemeLB (Network)              & AMD Opteron 6276 (9.2)   & 32      \\
HemeLB (Large Network)        & AMD Opteron 6276 (9.2)   & 512     \\
LB3Dv7 (Shamardin p.c.)        & AMD Opteron 6276 (9.2)   & 32      \\
LB3Dv7-3phase (Shamardin p.c.) & AMD Opteron 6276 (9.2)   & 128      \\
LBMHD~\cite{Williams:2011}     & AMD Opteron 1356 (9.2)   & 8192    \\
LBMHD~\cite{Williams:2011}     & AMD Opteron 6172 (8.4)   & 49152 \\
LUDWIG~\cite{Gray:2011}      & AMD Opteron 6276 (9.2)   & 384 \\
Palabos~\cite{Palaboswiki}   & AMD Opteron 8356 (9.2)   & 4       \\
\hline
HYPO4D (Groen p.c.)            & BlueGene/P (3.4)        & 512     \\
%LB3D (Schmieschek p.c.)        & BlueGene/P (3.4)        & 256     \\
LBMHD~\cite{Williams:2011}     & BlueGene/P (3.4)        & 8196       \\
Palabos~\cite{Palaboswiki}     & BlueGene/P (3.4)        & 256     \\
MUPHY~\cite{Bernaschi:2009}    & BlueGene/L (2.8)         & 32      \\
OpenLB~\cite{Fietz:2012}       & Intel Xeon X5355 (10.64) & 8       \\
Palabos~\cite{Palaboswiki}     & Intel Xeon X5550 (10.64) & 4       \\
HemeLB (Bifurcation, Sect~\ref{Sect:supermuc}) & Xeon E5-2680 (21.6)      & 128      \\
\hline
\end{tabular}
\end{table*}

\begin{table*}[!t]

\caption{Performance comparison of 12 LB simulations in our code comparison.
We provide the name of the LB application used in the first column, followed by
the number of lattice sites for each run, the directionality, and the obtained
performance per core. We give the per core calculation performance in millions
of site updates per second (MSUPS). In the case of LBMHD we assumed 1300 FLOPs
per lattice operation, as mentioned in Williams {\em et
al.}~\cite{Williams:2009, Williams:2011}. Runs that use a sparse simulation
domain are marked with an asterisk. Three-phase flow runs requires considerably 
more FLOPs per site update than single-phase flow runs.
Here, the OpenLB run used a data set with a fluid fraction of 0.145.  The
Palabos run on the Opteron relied on shared memory and multi-threading, and did
not use MPI.}

\label{Tab:CodeComparison2}
\centering
\begin{tabular}{|llll|}
\hline
Name & \# of lattice sites & directional & MSUPS\\
     &                     & resolution  & per core\\
\hline
HemeLB  (Cylinder)        & 15,607,040      & D3Q15 & 1.41\\
HemeLB* (Network)         & 18,836,545      & D3Q15 & 1.20\\
HemeLB* (Large Network)   & 44,650,496      & D3Q15 & 1.19\\
LB3Dv7                    & 16,777,216      & D3Q19 & 0.30\\
LB3Dv7 (3-phase flow)     & 56,623,104      & D3Q19 & 0.084\\
LBMHD (w/ magnetism)      & 6,115,295,232   & D3Q27 & $\sim1.42$\\
LBMHD (w/ magnetism)      & 28,311,552,000  & D3Q27 & $\sim1.15$\\
LUDWIG                    & 339,738,624     & D3Q19 & $\sim3.0$\\
Palabos (shared memory)   & 64,481,201      & D3Q19 & 2.55\\
\hline
HYPO4D  & 452,984,832     & D3Q19 & 0.273\\
%LB3D    & 2,147,483,648   & D3Q19 & 0.172\\
LBMHD   & 1,811,939,328   & D3Q27 & $\sim0.5$\\
Palabos & 1,003,003,001   & D3Q19 & 0.891\\
LUDWIG  & 16,777,214      & D3Q19 & 0.087\\
MUPHY   & 262,144         & D3Q19 & 0.529\\
OpenLB* & 1,060,000       & D3Q19 & $\sim0.4$\\
Palabos & 64,481,201      & D3Q19 & 7.87\\
HemeLB* (Bifurcation) & 19,808,107 & D3Q15 & 3.49\\
\hline
\end{tabular}
\end{table*}

\section{Modelling the performance of HemeLB}

\begin{table}[!t]
\caption{List of constant values used in our performance model. The $\lambda$ value was measured using a {\tt ping} test between nodes on HECToR.
The $\sigma$ value was taken by dividing the MPI point-to-point bandwidth specification on the HECToR website~\cite{hector} (at least 5 GB/s) by the number of 
cores per node (32).}
\label{Tab:constants}
\centering
\begin{tabular}{|l|l|}
\hline
Constant name & Value\\
\hline
$\tau$                   & $1.57 \times 10^{6}$ SUPS per core (calc only)\\
$\lambda$                & $2.5 \times 10^{-5} [s]$\\
$\sigma$                 & 160 MB/s per core\\
$\zeta_{\rm calc}$       & $1.04$\\
$\zeta_{\rm comm}$       & $1.5$\\
$O_{\rm monitoring}$     & $0.06$\\
\hline
\end{tabular}
\end{table}

\subsection{Parameter extraction}\label{Sect:extract}

Before we are able to construct and apply the performance model, we need to
extract a number of parameters specific to HemeLB. These parameters include
the maximum neighbour count, the communication 
volume and the calculation and communication load imbalance.

\subsubsection{Characterising maximum neighbour count}

Each process within HemeLB (except for the steering process) models a 
subsection of the simulation domain, and exchanges information with its 
neighbours. Here we characterise the maximum neighbour count ($k_{\rm max}$),
which is an approximation of the maximum number of neighbours a process has
in a given simulation.

To obtain the neighbour counts of each process, we have run the initialisation
routine of HemeLB (without any simulation time steps) using 4 to 16384 cores.
The number of neighbours is dependent not only on core count but also on the
geometry of the simulation domain, which makes it non-trivial to fully
approximate it in the performance model. Instead, we choose to model close to a
worst-case decomposition scenario, selecting the simulation domain with the
highest neighbour count, and using the measured values there to determine
$k_{\rm max}$ for any simulation domain. Because ParMETIS does not guarantee a
reproducible decomposition, simulations may vary in neighbour counts for a 
given problem on a given number of cores. We therefore have repeated each 
measurement three times.

\begin{figure}[!t]
\centering
\includegraphics[width=5.0in]{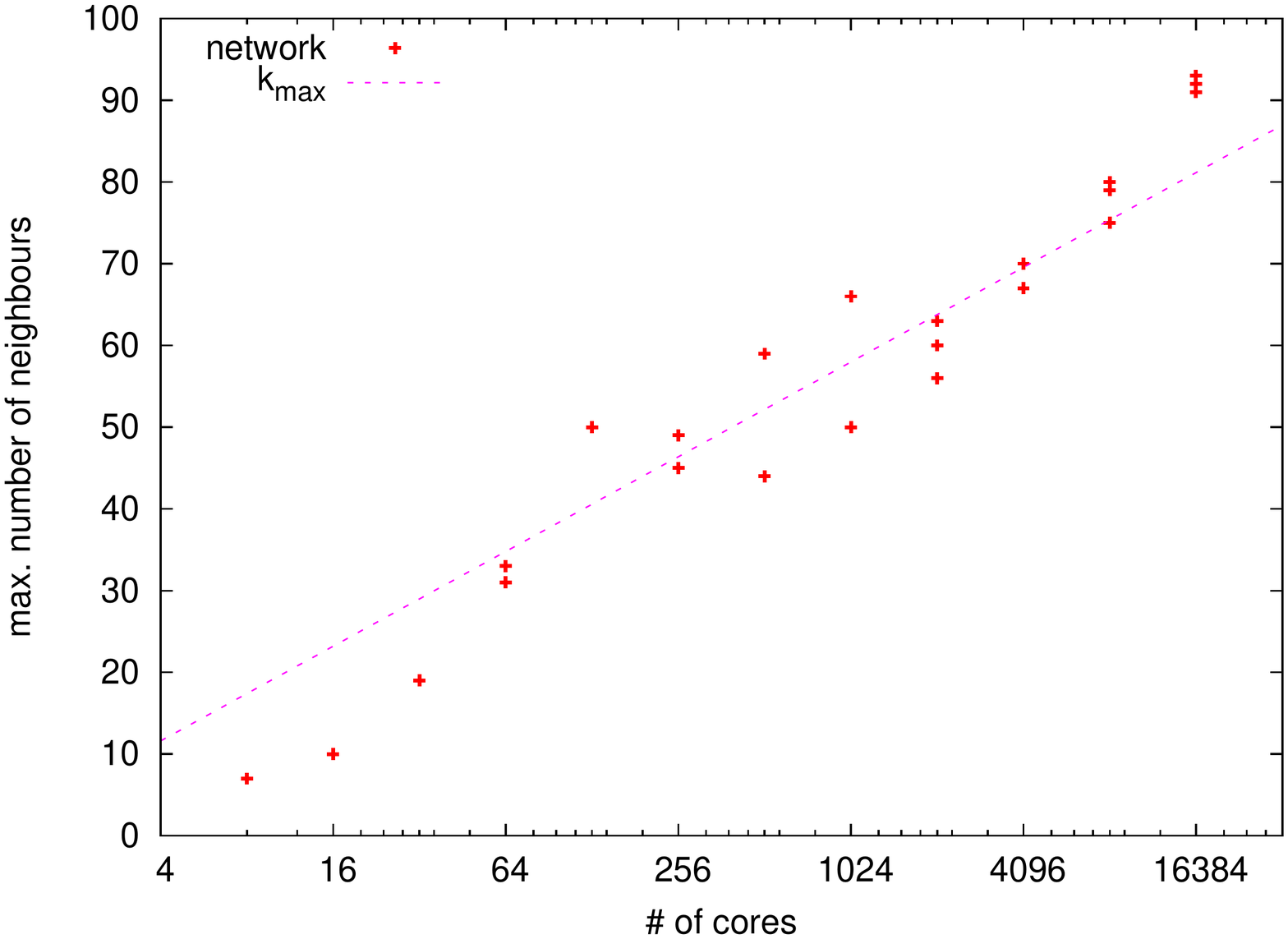}
\caption{Maximum number of neighbours as a function of the core count. Here
we selected and fitted our model to the Network simulation domain, which
has the highest neighbour count due to its sparseness. We ran 3 decomposition
routines for each core count included in the figure, plotting the highest
neighbour count seperately for each instance. The neighbour count
approximation used in our performance model is given by the dotted line.}
\label{Fig:neighbours}
\end{figure}

We present our measurements of the maximum neighbour count as a function of the
core count in Figure~\ref{Fig:neighbours}. We find that the maximum neighbour
count for the network geometry ranges from 7 on 8 cores, up to as high as 94 
on 16,384 cores. Based on this data, we created a logarithmic fit, approximating 
$k_{\rm max}$ as:

\begin{eqnarray}
  k_{\rm max} =  \frac{\log{p}}{\log{1.127}}.
\end{eqnarray}

\subsubsection{Characterising communication volume}

\begin{figure}[!t]
\centering
\includegraphics[width=5.0in]{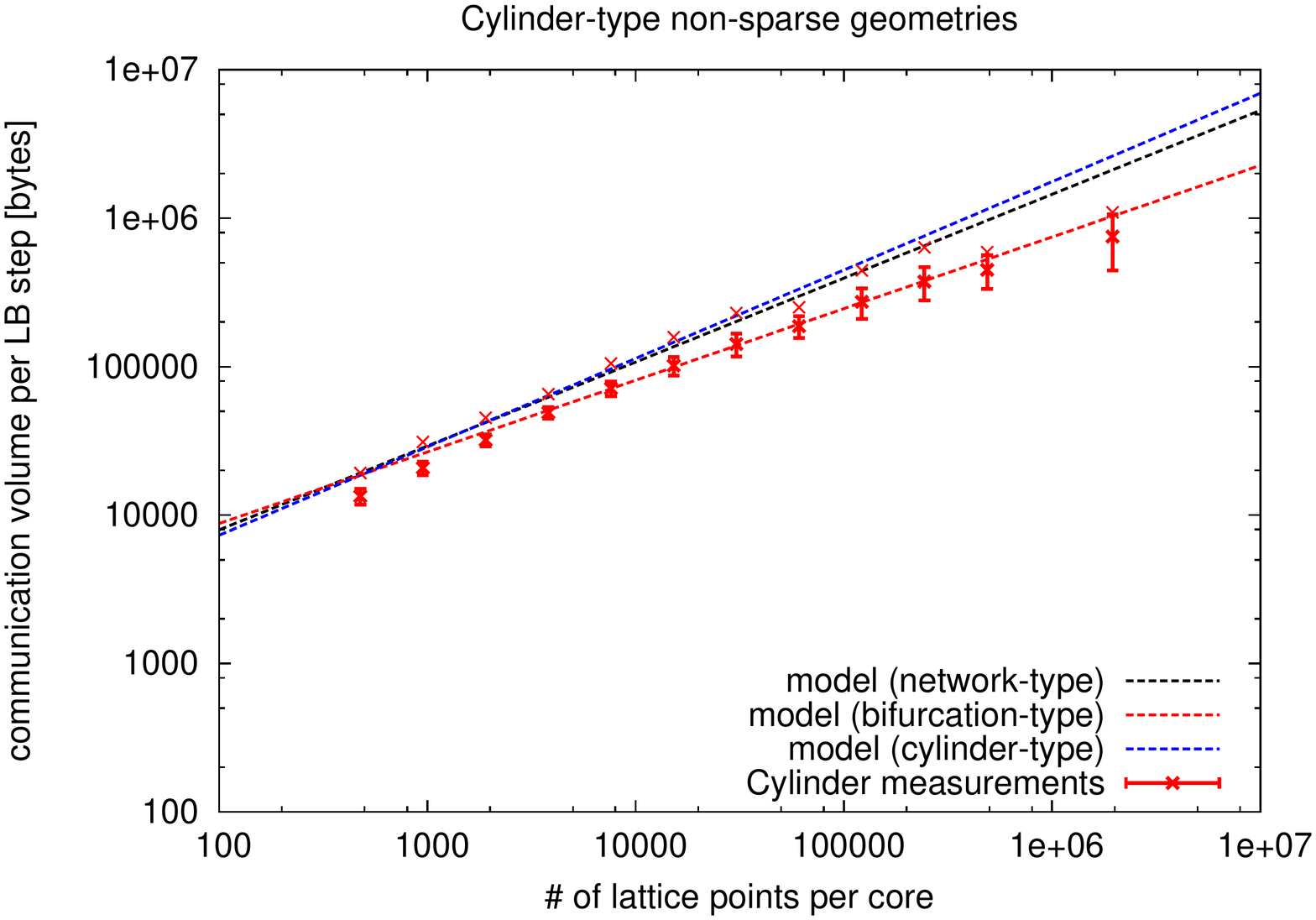}
\caption{Number of bytes sent per LB simulation step as a function of the number of lattice sites per core 
for Cylindrical geometries (measurements have been done using the Cylinder simulation domain). The fits we use
for Cylindrical, Bifurcation and Network geometries in our performance model are given respectively by the red, 
blue and black dashed lines. Error bars show one standard deviation for the distribution across cores.}
\label{Fig:C}
\end{figure}

\begin{figure}[!t]
\centering
\includegraphics[width=5.0in]{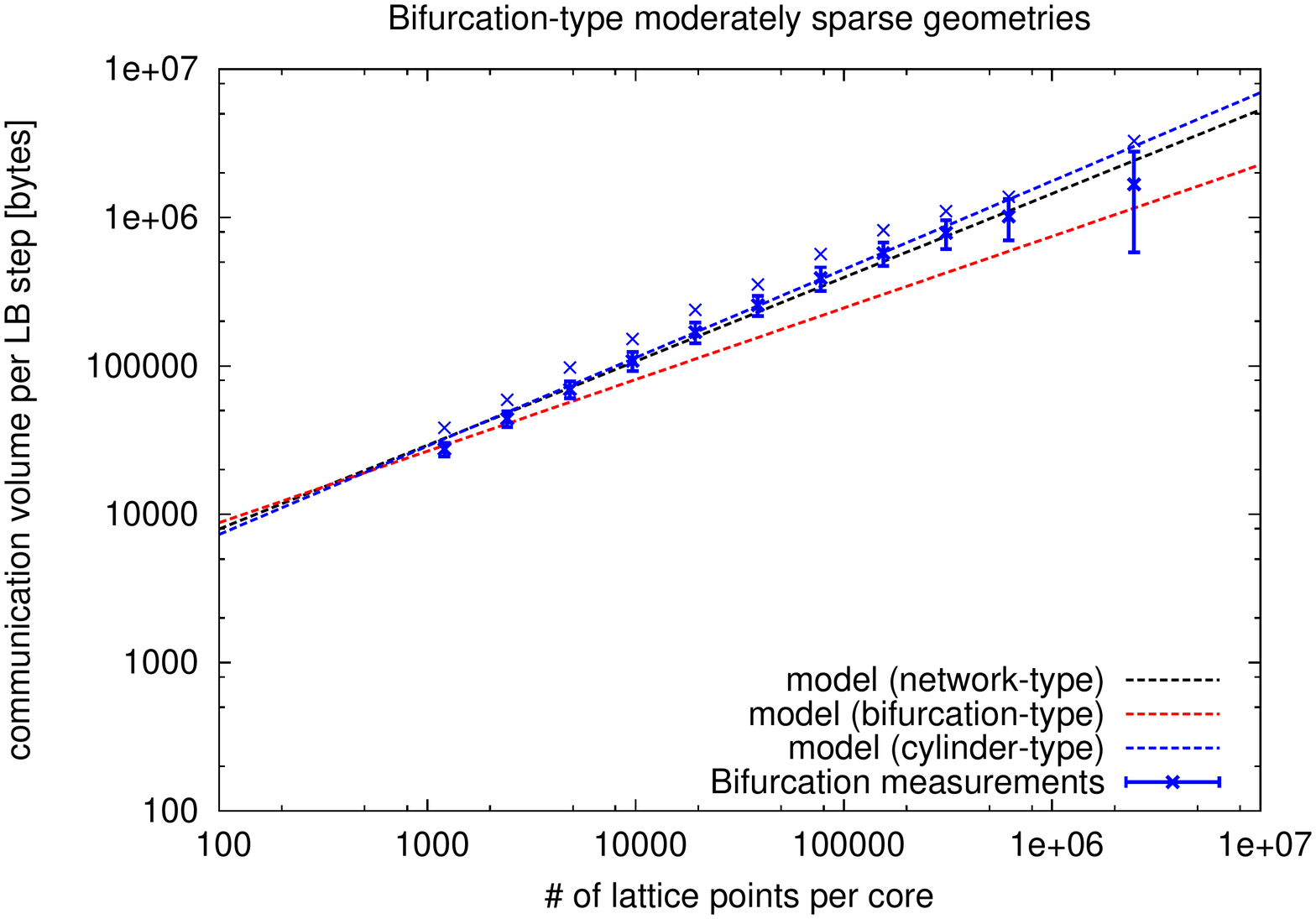}
\caption{As in Fig.~\ref{Fig:C} but for the Bifurcation geometry (using the Bifurcation simulation domain).}
\label{Fig:B}
\end{figure}

\begin{figure}[!t]
\centering
\includegraphics[width=5.0in]{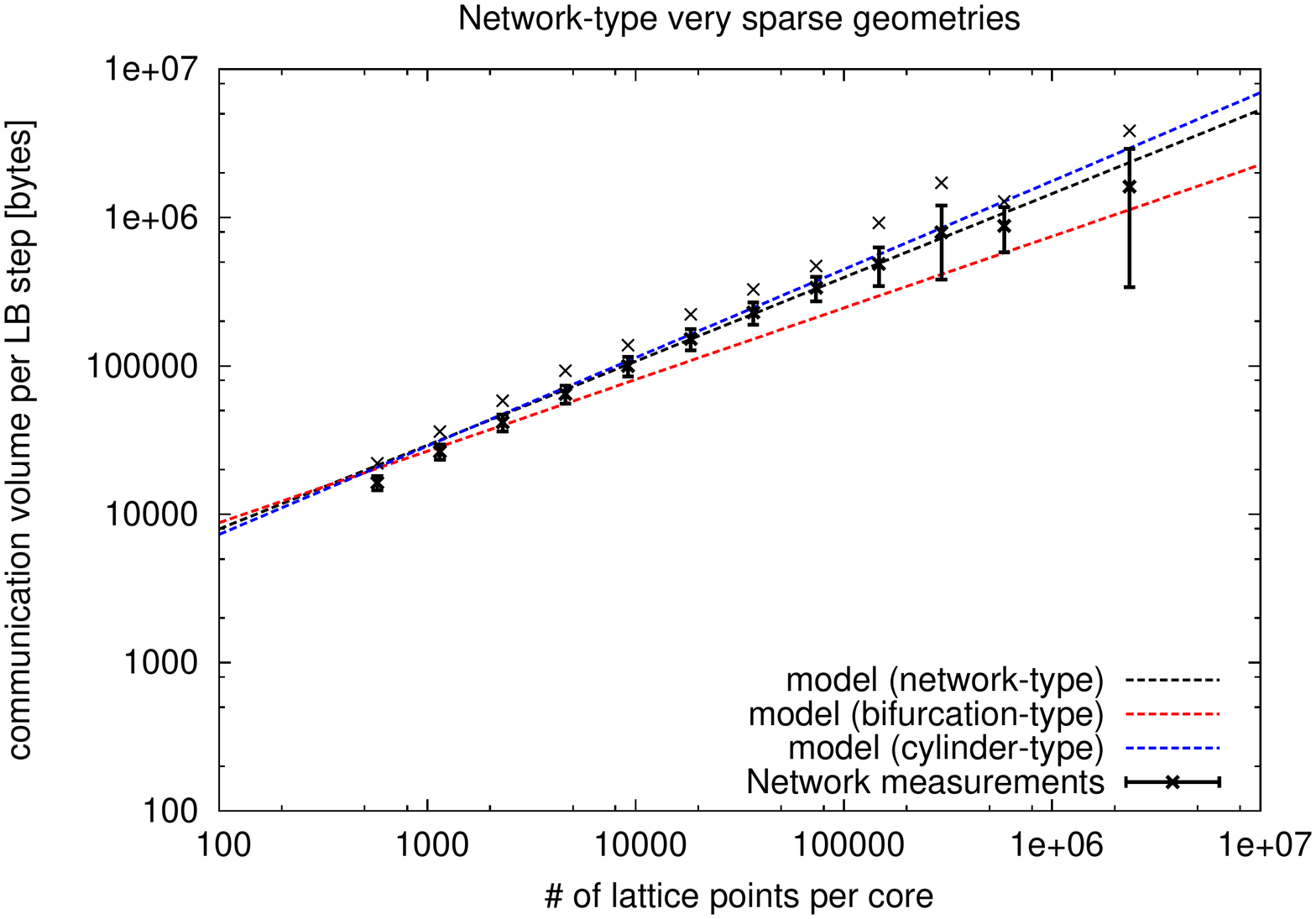}
\caption{As in Fig.~\ref{Fig:C} but for the Network geometry (using the Network simulation domain).}
\label{Fig:N}
\end{figure}

To model the communication performance of HemeLB we also need information on the
amount of data communicated between processes at each step. As the domain
decomposition in HemeLB is done at runtime~\cite{Mazzeo:2008}, we can only know the exact
communication data volume after we have launched the simulation. To preserve
the predictive power of the performance model, we have instead measured the
communication volume for the three types of simulation domains across a range
of core counts. After having performed the measurements, we fitted the data
to a function of the form $a x^b$ to gain an approximate estimate while 
keeping the model relatively straightforward.
We present our measurements of the communication volume and our fits for the
cylindrical geometries in Figure~\ref{Fig:C}, for the bifurcation geometries in
Figure~\ref{Fig:B}, and for the network geometries in Figure~\ref{Fig:N}. Here
we find that the communication volume can differ by as much as a factor four
between the domain types, making separate fits necessary for each type. We
provide the exact formulation for each of the three fits in Table~\ref{Tab:Fits}. 
Interestingly, these scale with less than $(N/P)^{2/3}$ as one 
would expect with, for example, a decomposition into cubes. At large N/p, i.e. few 
processes, the sparseness implies that large parts of the single-process volumes
are bordered by boundary sites, rather than lattice sites residing on neighbouring
processes. We therefore observe a scaling of less than $(N/P)^{2/3}$. 
In the limit of small (N/P), the measured communication volume does converge 
to the function $S=250 \times (N/P)^{2/3}$ when the number of sites per process 
becomes lower, and the number of cores used higher in the simulations. Because
the process-specific volumes are smaller here, the sparseness of the domain has
a smaller effect on the measured (maximum) neighbour count.

\subsubsection{Characterising load imbalances} 
When using sparse geometries, individual processes within HemeLB contain
subsets of the simulated system with heterogeneous shapes and sizes. These
differences result in two types of load imbalance during the parallel LB
calculation: a calculation load imbalance and a communication load imbalance.
To obtain a platform-independent measure of the load imbalance in HemeLB, we
choose not to include timing results in this procedure. Instead, we examine the
number of lattice sites on each core to determine the calculation imbalance and
the number of bytes sent by each process to determine the communication
imbalance. Both metrics are reproducible on different platforms when using the
same version of ParMETIS (4.0.2), although some variations may occur due to
the stochastic nature of the ParMETIS decomposition technique.

\begin{figure}[!t]
\centering
\includegraphics[width=5.0in]{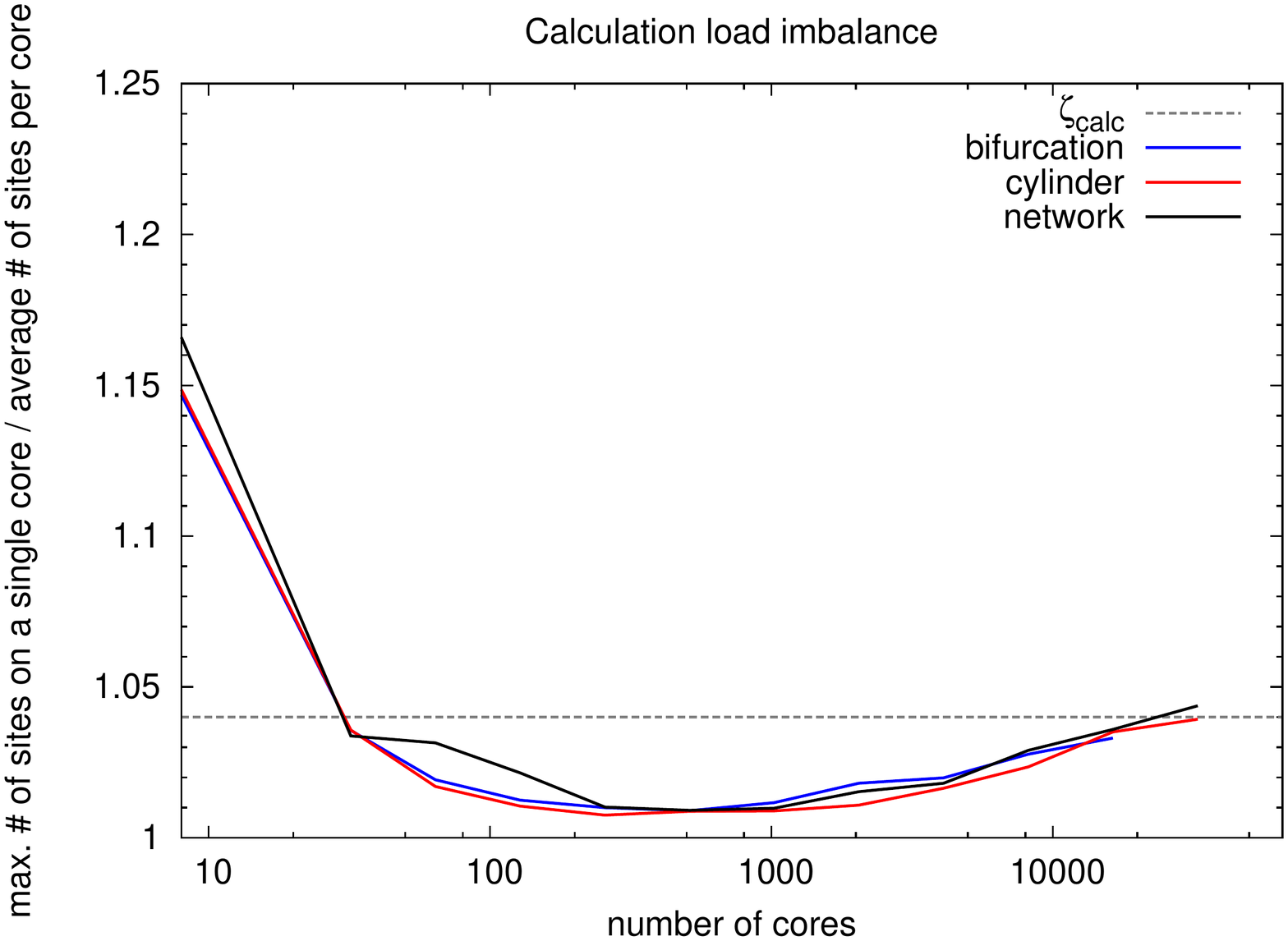}
\caption{Imbalance of the number of sites per core (i.e., our measure for calculation load imbalance) as a function of the number of cores for the three geometries. The value on the y-axis is the relative calculation overhead caused by load imbalance. These values are deterministic for a given core count and ParMETIS version.}
\label{Fig:CalcBalance}
\end{figure}

In Fig.~\ref{Fig:CalcBalance} we show the measured {\em calculation load
imbalance} for three geometries as a function of the core count. We
determine this calculation load imbalance by dividing the maximum number of
lattice sites on any core within this run by the average number of sites over
all cores in the same run. HemeLB is optimised for calculation load balance and we find an
imbalance of less than $1.04$ for most core counts. However, the calculation
load imbalance is higher for both very low and very high core counts. This 
contributes in part to the superlinear scaling of HemeLB at lower
core counts in some cases, and reduces scalability when there are less
than 2000 lattice sites per core. Based on these measurements, we assume a
calculation load imbalance ($\zeta_{\rm calc}$) of 1.04 in our performance model.
 
\begin{figure}[!t]
\centering
\includegraphics[width=5.0in]{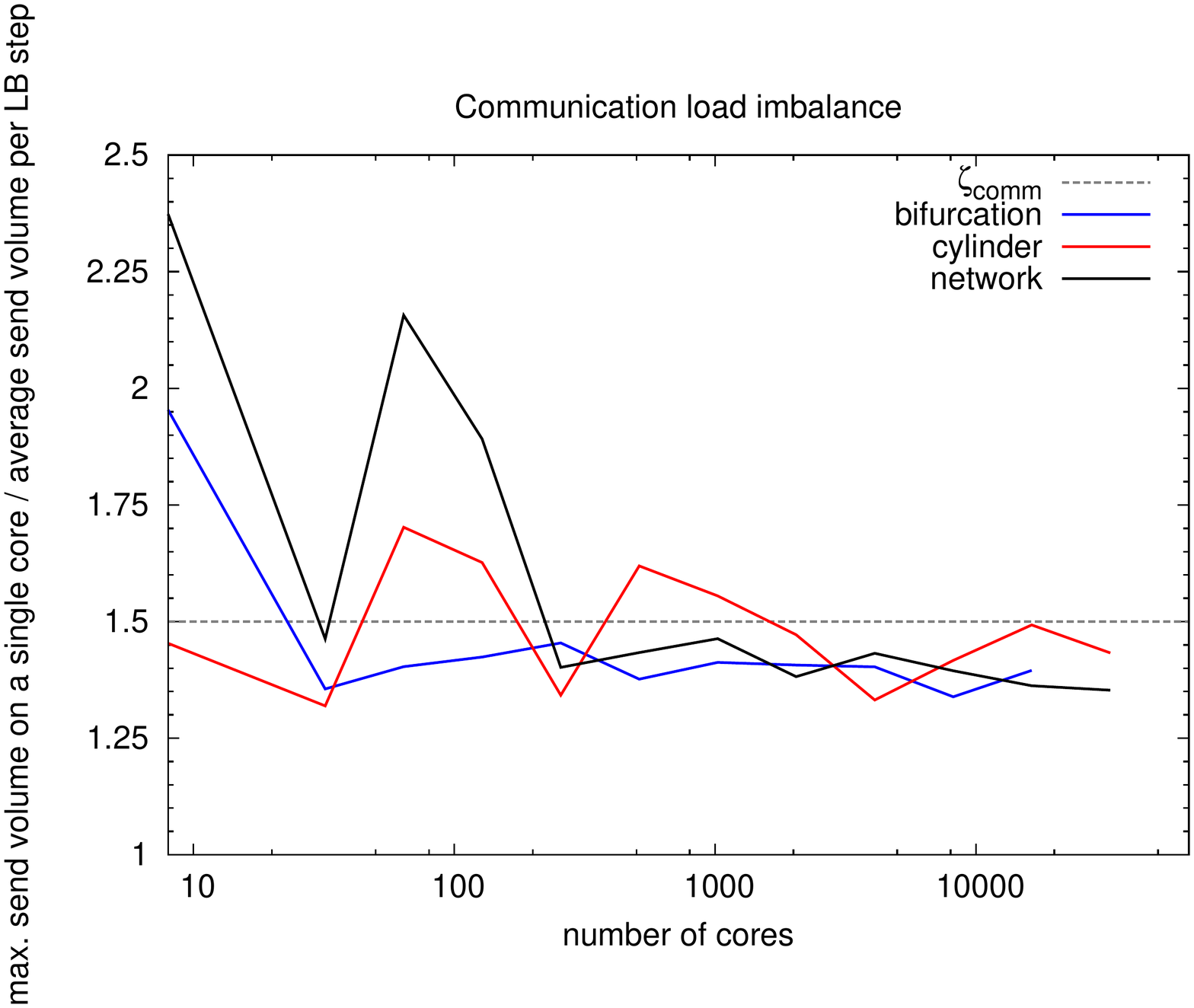}
\caption{Imbalance in the number of bytes sent (i.e., our measure for communication load imbalance) as a function of the number of cores for the three geometries. The value on the y-axis is the relative communication overhead caused by load imbalance. These values are deterministic for a given core count and ParMETIS version.}
\label{Fig:CommBalance}
\end{figure}

In Fig.~\ref{Fig:CommBalance} we present the {\em communication load
imbalance}, which we measure by dividing the maximum number of bytes sent by a
single core in the run by the average number of bytes sent per core. All the
communication measurements are given per step. We observe a large and erratic
imbalance in the communication sizes. The ParMETIS domain distribution
algorithm co-optimizes for both calculation load balance and communication
minimisation.  However, these results suggest that it does not optimize for
communication balance. This communication imbalance does not strongly diminish
the code performance unless the performance is already dominated by
communication. Within our model we take an approximate average of our
measurements, and assume a communication load imbalance ($\zeta_{\rm comm}$) of
1.5.

\begin{table}[!t]
\caption{List of fitting functions used in our performance model. Here the total number 
of lattice sites is given by $N$ and the number of cores used by $p$.}
\label{Tab:Fits}
\centering
\begin{tabular}{|ll|}
\hline
Constant name & Value\\
$S_{\rm cylinder}$    & $1898  \times (N/p)^{0.482719}$ bytes per core per step\\
$S_{\rm bifurcation}$ & $942.0 \times (N/p)^{0.595517}$ bytes per core per step\\
$S_{\rm network}$     & $1176  \times (N/p)^{0.613449}$ bytes per core per step\\
\hline
\end{tabular}
\end{table}

\subsection{LB calculations}

To model the performance of the core LB simulator code we propose a time-complexity 
model which is loosely based on \cite{Axner:2007,Axner:2008} but largely simplified. We use a range of 
parameters which we derived in Section~\ref{Sect:extract}. In this model we approximate the 
overall time spent to perform a single simulation step in HemeLB ($T_{\rm step}$), using

\begin{eqnarray}
  T_{\rm step} =  \frac{\zeta_{\rm calc} \times T_{\rm calc} + \zeta_{\rm comm} \times T_{\rm comm}}{1.0 - O_{\rm monitoring}},
\end{eqnarray}
where $T_{\rm calc}$ is the average calculation time per core, ($\zeta_{\rm calc}$) is
the calculation load imbalance constant, $T_{\rm comm}$ is the communication time per core, $\zeta_{\rm comm}$ is the communication load 
imbalance constant and $O_{\rm monitoring}$ is the fraction
of time spent on monitoring overhead. Throughout our runs we
found that $\sim6\%$ of the runtime is spent on monitoring, so we define
$O_{\rm monitoring} = 0.06$. The average calculation time per core is 
given by
\begin{eqnarray}
  T_{\rm calc} = \frac{\left(N/p\right)}{\tau}
\end{eqnarray}
Here, the total number of lattice sites is given by $N$ and the number of cores
by $p$.  We define the SUPS per core $\tau$
as a platform dependent constant for the HECToR machine in Table~\ref{Tab:constants}. 
We measured $\tau$ as an average from our HemeLB runs with 32 cores (1 node). The 
true SUPS capacity per core depends slightly on the number of
sites per core, but is in almost all cases within $20\%$ of this average value.
We model the time spent on communications, $T_{\rm comm}$, using

\begin{eqnarray}
  T_{\rm comm} = \log_2(p) \times \lambda + \frac{S_{\rm <x>}}{\sigma},
\end{eqnarray}
where $\lambda$ is the point-to-point latency of MPI communications between nodes in seconds, 
and $\sigma$ the average throughput capacity per core in bytes. We assume that the number of messages exchanged per time step increases with the number of processes and we model this as $\log(p)$. The number of bytes sent out 
per core per step ($S_{\rm <x>}$) is dependent on the geometry used as well as the number of sites per core. 
We have provided basic fits for three geometry layouts with different sparsity (network, bifurcation 
and cylinder) in Table~\ref{Tab:constants}. These fits are most accurate for simulations that
have between 5,000 and 200,000 sites per core.

\subsection{Visualisation}\label{Sect:vis-model}

When image rendering and writing is enabled in HemeLB, some overhead is introduced
in the execution, and the new time per step ($T_{\rm step\_vis}$) becomes

\begin{eqnarray}
T_{\rm step\_vis} = T_{\rm step} + T_{\rm images},
\end{eqnarray}
where $T_{\rm images}$ is the overhead for rendering and writing images. Because our
overhead measurements show a large variability, we use a straightforward fit rather
than a detailed sub-model to approximate this overhead. Based on our measurements on
2048 cores, we have derived an approximate fit of $T_{\rm images} = 21.6 k^{-0.76}$, with 
$k$ being the number of LB steps per rendered image. We provide a graphical overview of 
the approximation in Figure~\ref{Fig:Scalability-vis-render}.

%\begin{figure}[!t]
%\centering
%\includegraphics[width=5.0in]{scalability-vis-render-validate.pdf}
%\caption{Overhead in seconds relative to the simulation time without images
%rendered as a function of the number of LB steps per image rendered and written.
%Error bars are present in Figure~\ref{Fig:Scalability-vis-render}, but omitted 
%here for clarity. The prediction of the performance model is given by the thick 
%solid red curve.}
%\label{Fig:Scalability-vis-model}
%\end{figure}

\section{Model validation}

\subsection{Validation on HECToR}
\begin{figure}[!t] \centering
\includegraphics[width=5.0in]{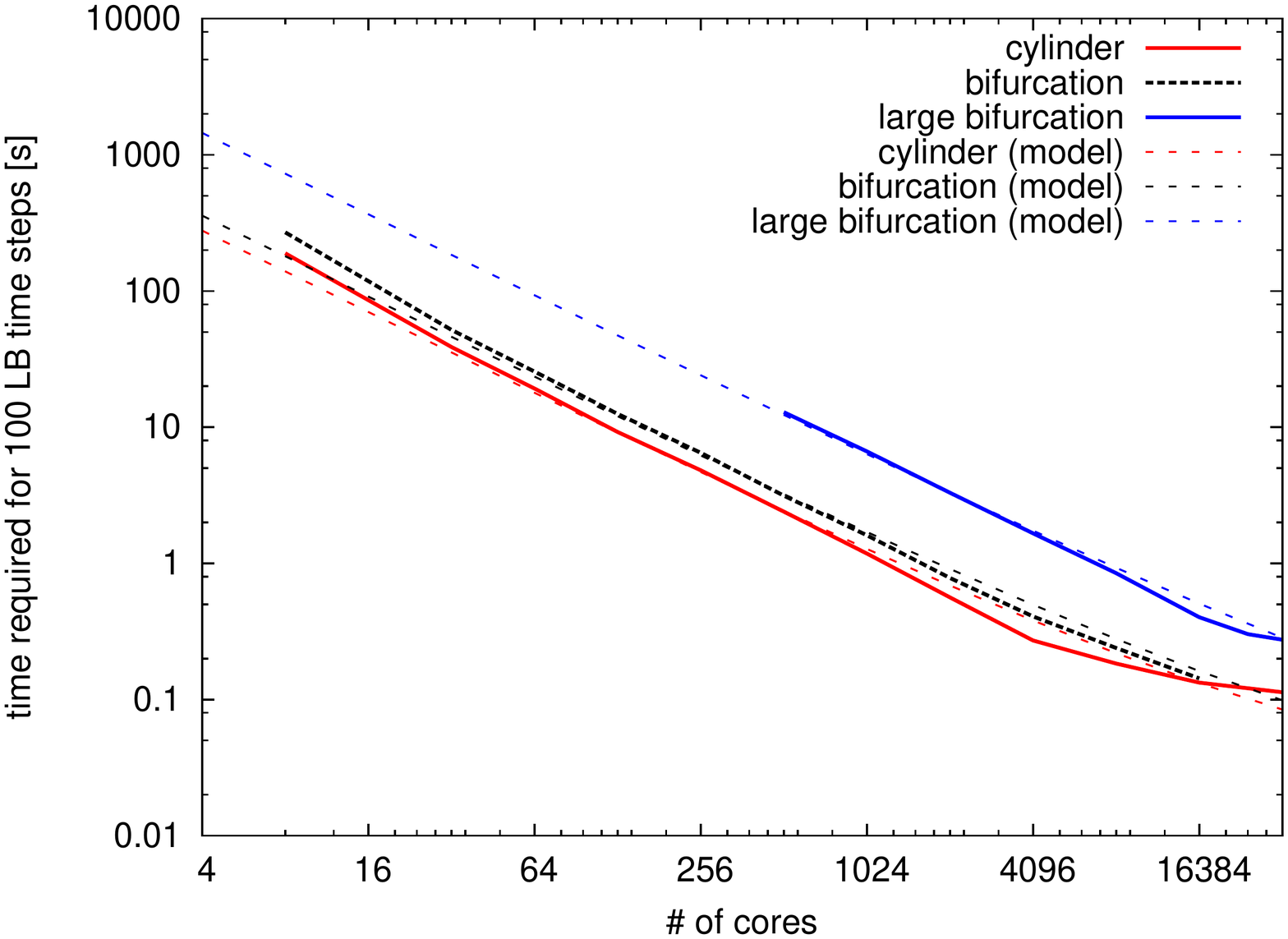}
\caption{Wall-clock time spent to simulate 100 time steps as a function of the
number of cores used for the Cylinder, Bifurcation and Large Bifurcation
simulation domains. These validation runs were done using the HECToR
supercomputer. Predictions by our performance model are indicated by the dashed
lines.} \label{Fig:Scalability} \end{figure}

\begin{figure}[!t]
\centering
\includegraphics[width=5.0in]{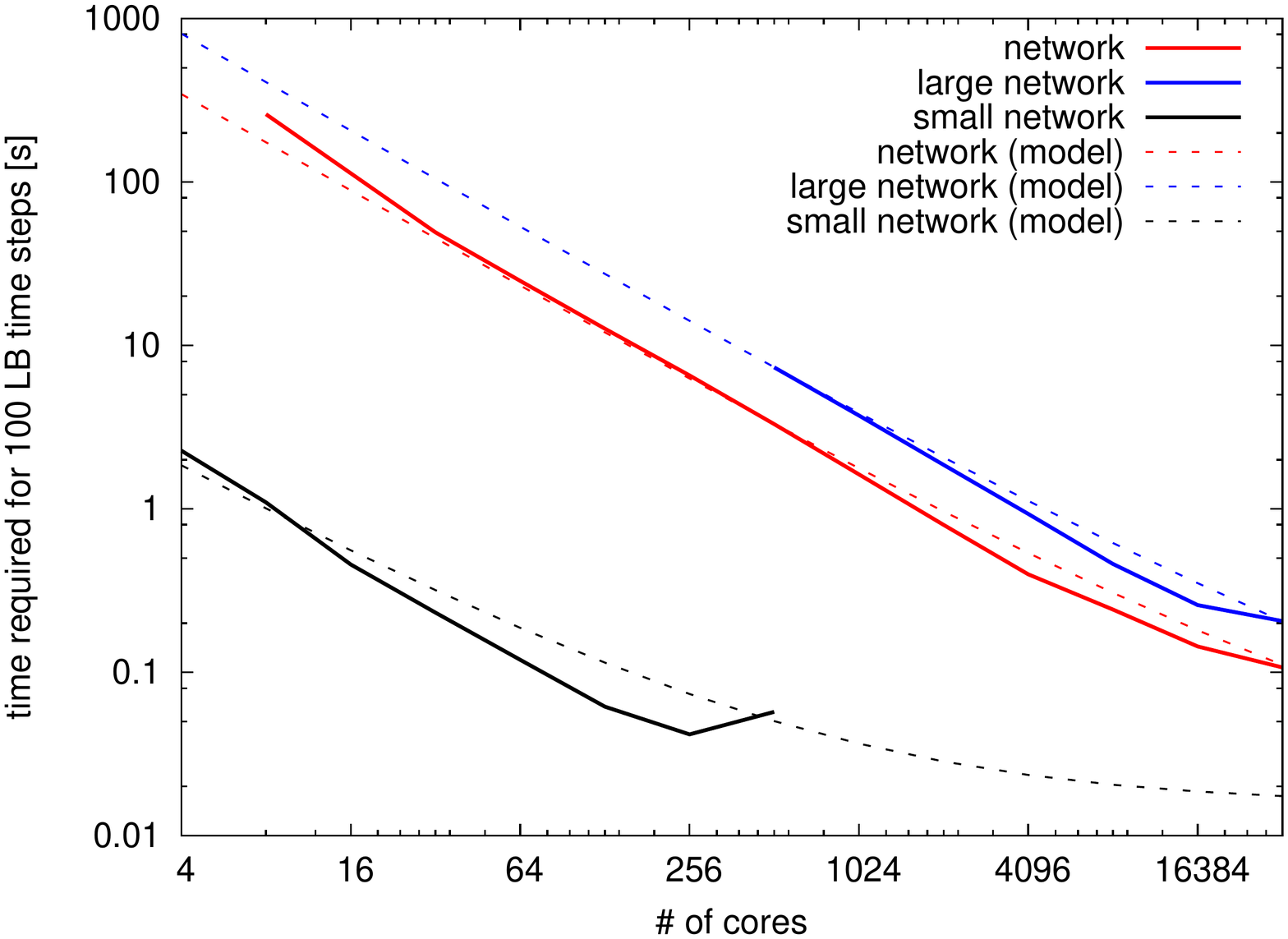}
\caption{As Fig.~\ref{Fig:Scalability}, but for the Network, Small Network and Large Network simulation domains. These validation runs were done using the HECToR supercomputer.}
\label{Fig:Scalability-network}
\end{figure}

We have applied our performance model to calculate the theoretical execution times
of the simulations we presented in Section~\ref{Sect:coreLBperf}. The
predictions given by the model, as well as the measurements presented earlier,
can be found in Figure~\ref{Fig:Scalability} for the Cylinder, Bifurcation and
Large Bifurcation simulation domains and in
Figure~\ref{Fig:Scalability-network} for the Network, Small Network and Large
Network simulation domains. The predictions from our model are generally in
agreement with our measurements, especially for the larger simulation domains.
However, the model does not reproduce the superlinear speedup measured in the
results. This is mainly because the model assumes a constant calculation and 
communication load imbalance, regardless of core count. In contract we measure 
relatively large calculation and communication load imbalances for runs on
less than 32 cores (see Figures~\ref{Fig:CalcBalance} and \ref{Fig:CommBalance}). 
In this regime, the measured load imbalances are considerably higher than the 
ones assumed in our model, and the execution time is consequently slightly higher
than in our model predictions.

\subsection{Validation on SuperMUC}~\label{Sect:supermuc}

To test whether our performance model holds when applied to a different
platform, we used a small part of an allocation arranged by MAPPER on the
SuperMUC supercomputer at the Leibniz-Rechenzentrum in Garching, Germany.
SuperMUC is an IBM System x iDataPlex machine with 147456 compute cores and a
total peak performance of 3.185 PFLOP/s (21.6 GFLOP/s per core). Each node has 16 cores,
consists of two Intel Xeon E5-2680 CPUs, and is equipped with 32 GB of memory.
The nodes are interconnected with an Infiniband FDR10 network, which divides
the supercomputer into {\em islands}, each of which contains 8192 cores.  We
use this machine to run HemeLB simulations using the Bifurcation simulation
domain, and to compare our measurements to our model predictions. We provide
the list of constant values in Table~\ref{Tab:constantsMUC}. The values of
$\zeta_{\rm calc}$, $\zeta_{\rm comm}$ and $O_{\rm monitoring}$ are the same as
those we used for HECToR, as these constants do not depend on the underlying
architecture. We obtained a $\sigma$ value of 500 MB/s per core through direct
correspondence with LRZ, and measured a $\lambda$ of $1.83 \times 10^{-4}$ s by
running a ping job between two nodes within the same island, and taking the
average from 10 pings. As all small jobs on SuperMUC tend to get scheduled on
the same island, it was unfortunately not possible to accurately measure the
latency between islands. We obtained the value of $\tau$ by running a very
short HemeLB simulation on one node and extracting the calculation rate per core,
excluding any communications or other overhead ($4.2 \times 10^{6}$ SUPS).

\begin{table}[!t]
\caption{List of constant values used in our performance model for SuperMUC. The $\lambda$ value was measured using a {\tt ping} test between nodes on SuperMUC.
The $\sigma$ value was taken by dividing the MPI point-to-point bandwidth specification on the SuperMUC website~\cite{hector} (at least 5 GB/s) by the number of
cores per node (16).}
\label{Tab:constantsMUC}
\centering
\begin{tabular}{|l|l|}
\hline
Constant name & Value\\
\hline
$\tau$                   & $4.2 \times 10^{6}$ SUPS per core (calc only)\\
$\lambda$                & $1.83 \times 10^{-4} [s]$\\
$\sigma$                 & 500 MB/s per core\\
\hline
\end{tabular}
\end{table}

\begin{figure}[!t]
\centering
\includegraphics[width=5.0in]{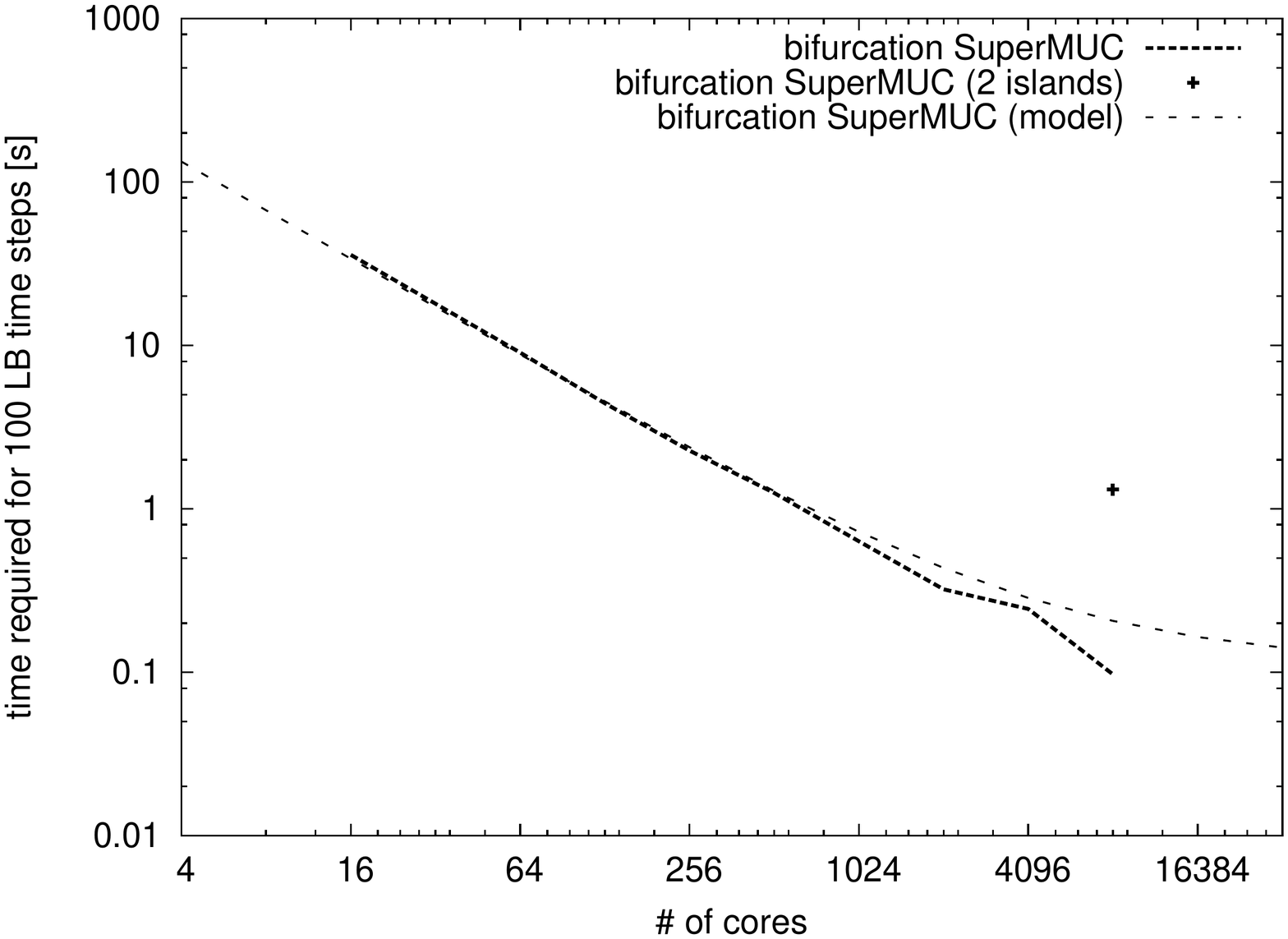}
\caption{Wall-clock time spent to simulate 100 time steps as a function of the number of 
cores used for the Bifurcation simulation domains, using the SuperMUC supercomputer. Predictions by our performance model are indicated by the dashed line.}
\label{Fig:Scalability-supermuc}
\end{figure}

We present both our model predictions and our performance measurements in
Figure~\ref{Fig:Scalability-supermuc}. Here we find that our simulation runs
considerably faster on SuperMUC than on HECToR, achieving 3.49 MSUPS per core
when using 128 cores, and 3.00 MSUPS per core when using 2048 cores. Our
performance model accurately predicts the runtime for simulations up to 4096
cores, and matches the measured performance even more closely than in the
HECToR validation tests. We have performed 2 runs using 8192 cores, one using
one island, and on distributed over two islands. Our performance model predicts
a time which is higher than the measured time for the single-island run. This 
may be because the {\tt ping} test we used over the 
Infiniband has given us a somewhat higher $\lambda$ than the actual 
point-to-point latency of communications in the MPI layer. Communications between islands 
experience much higher latency and lower bandwidth (at a 4:1 ratio). As a result, 
the run performed using two islands is an order of magnitude slower slower than the run
using one island. Understanding the performance across islands would require
us to assess the latency and bandwidth characteristics of the inter-island 
links (which would require a special access mode), and incorporate these in 
a separate "inter-island" set of the parameters $\lambda$ and $\sigma$.

\section{Discussion}

We have presented a range of performance measurements for HemeLB, covering the
lattice Boltzmann simulation and the visualisation and steering
functionalities. For the models studied here, HemeLB scales near-linearly up to
32,768 cores, even for highly sparse simulation domains such as vascular
networks. The application achieves close to maximum efficiency when using
between 5,000 and 500,000 lattice sites per core. We have shown that HemeLB can
render and write images once every 100 timesteps with an overhead of $\sim$10\%,
sharing streaming images and control with a steering client at 4.6 fps with a
28\% overhead.  We have demonstrated that it is possible to create a model
which can estimate the run time of HemeLB simulations in advance. In our
validation tests, we find that the predictions are between 70\% and 140\% of
the actual runtime for simulations with at least 5000 lattice sites per core,
and that our model remains largely accurate when applied to a different
architecture (SuperMUC).  We believe that accurate runtime predictions will be
useful in the long term when HemeLB is used in a clinical setting, as doctors
will be able to select the simulation with the highest accuracy that still
meets the deadline for actual treatment. 

To improve the accuracy of HemeLB simulations, as part of the MAPPER
project~\cite{MAPPER}, we have developed an intercommunication layer that
allows the code to exchange boundary information with other simulation
codes~\cite{Groen:2012-2}. These couplings allow us to
incorporate phenomena that are not resolved in HemeLB itself, such as the
interaction between the blood flow in the intracranial vasculature and that in
the rest of the human body. The boundary exchanges in these coupled simulations
occur at high frequency and require rapid response times on both ends. The
performance bottlenecks we have identified allow us to take the necessary steps
to ensure an optimal performance for multiscale simulations using HemeLB.

The envisaged use-case for HemeLB, involving deployment within a clinical
setting, is made more difficult by typical queueing and scheduling policies for
supercomputers. One important benefit of supercomputing lies in enabling
results to be produced in a timely fashion. With typical scheduling policies,
however, many codes produce results only after a lengthy wait in a queuing
system, significantly reducing the value-added of the supercomputing resource
relative to a long-running simulation on a smaller machine. The value of
supercomputing is particularly apparent when using interactive visualisation
and steering~\cite{Mazzeo:2010}, as this enables complex simulations to be
investigated on timescales close to those of human engagement. However, this
form of interaction is not possible without an advance reservation facility,
enabling one to predict the time when one will be able to interact with the
running simulation. 

In particular, in the clinical context, patients and physicians already
interact within a complex resource availability and scheduling environment. In
this case, advance reservation will be necessary to make computing resources
available concurrently with medical equipment, physicians, and patient needs.
Furthermore, when HemeLB is used in a clinical context, rapid access to
computing resources will become a safety-critical factor. This requires not
just advance reservation, but support for urgent computing~\cite{Sadiq:2008}.
For the use cases we envisage for HemeLB, an urgent computing mechanism will
need to be available on supercomputing resources.

%Having assessed the performance of HemeLB, we are working to prepare HemeLB for
%more sophisticated scenarios. We are carrying out a comprehensive correctness
%review for HemeLB, with special attention to ensuring appropriateness of
%boundary conditions for physiologically realistic scenarios, looking towards
%future clinical validation.  We are coupling HemeLB to other, more
%coarse-grained blood flow codes within MAPPER, and are also preparing HemeLB
%simulations that include bio-colloids. Additionally, we aim to improve the
%HemeLB performance still further. We will focus on performance for high-quality
%remote rendering and on single core efficiency. As part of the
%CRESTA~\cite{CRESTA} project we will further improve the scaling to ready the
%code for use on exascale machines.

\section*{Acknowledgments}

We thank Dr. Lev Shamardin for his assistance in including the LB3D version 7
performance measurements and Dr. Timm Krueger for his valuable comments. Our
research has received funding from the CRESTA and MAPPER projects within the
European Community's Seventh Framework Programme (ICT-2011.9.13) under Grant
Agreements nos. 287703 and 261507, and by EPSRC Grants EP/I017909/1
(www.2020science.net) and EP/I034602/1. This work made use of HECToR, the UK's
national high-performance computing service, which is hosted by UoE HPCx Ltd at
the University of Edinburgh, Cray Inc and NAG Ltd, and funded by the Office of
Science and Technology through EPSRC's High End Computing Programme. In
addition we have made use of the SuperMUC supercomputer, hosted by the Leibniz
Rechenzentrum (LRZ) in Garching, Germany. We are grateful to the MAPPER
consortium for providing an allocation on SuperMUC.



\bibliographystyle{elsarticle-num}
\bibliography{Library}

\end{document}